\documentclass[12pt
]{article}
\usepackage{amssymb}
\def\inh{\vskip 0.075truein \noindent\hangindent=12 pt \hangafter=1}

\usepackage{amsthm}\theoremstyle{remark}
\usepackage{graphicx}
\usepackage{caption}
\usepackage{subcaption}
\usepackage[10pt]{type1ec}
\usepackage[T1]{fontenc}
\usepackage[polutonikogreek,english]{babel}

\newcommand{\bte}{\begin{quote}\begin{theorem}}
\newcommand{\ete}[1]{\label{#1}\end{theorem}\end{quote}}
\newcommand{\bcom}{\begin{quote}\end{quote}}
\newcommand{\bex}{\begin{quote}\begin{example}}
\newcommand{\eex}[1]{\label{#1}\end{example}\end{quote}}
\newcommand{\bcon}{\begin{quote}\begin{conclusion}}
\newcommand{\econ}[1]{\label{#1}\end{conclusion}\end{quote}}
\newcommand{\bdefi}{\begin{quote}\begin{definition}}
\newcommand{\edefi}[1]{\label{#1}\end{definition}\end{quote}}

\newcommand{\blem}{\begin{quote}\begin{lemma}}
\newcommand{\elem}[1]{\label{#1}\end{lemma}\end{quote}}

\newcommand{\bpr}{\begin{quote}\begin{problem}}
\newcommand{\epr}[1]{\label{#1}\end{problem}\end{quote}}

\newcommand{\beq}{\begin{eqnarray}}
\newcommand{\eeq}[1]{\label{#1}\end{eqnarray}}
\newcommand\eq[1]{(\ref{#1})}
\newcommand{\bfi}{\begin{figure}[24]}
\newcommand{\efi}[1]{\caption{\label{#1}}\end{figure}}
\newcommand\fig[1]{Fig.~\ref{#1}}

\newcommand{\bexe}{\begin{quote}\begin{exercise}\inh}
\newcommand{\eexe}[1]{\label{#1}\end{exercise}\end{quote}}

\def\AEM{ Adv. Eng. Mater.}

\def\IJSS{ Int. J. Solids Struct.}

\def\JE{ J. Elasticity}

\def\JEMT{ J. Eng. Mat. Tech.}

\def\JMS{ J. Mater. Sci.}
\def\JMPS{ J. Mech. Phys. Solids}

\def\JSV{ J. Sound Vib.}

\def\MOM{ Mech. Mater.}

\def\NJP{ New J. Phys.}

\def\PNAS{ Proc. Natl. Acad. Sci. USA}

\def\PRL{ Phys. Rev. Lett.}
\def\PRSLA{ Proc. R. Soc. Lond. A}

\def\PSSB{ Phys. Status Solidi B}

\def\QJMAM{ Quart. J. Mech. Appl. Math.}
\def\SMS{ Smart Mater. Struct.}

\usepackage{graphics,graphicx}
\usepackage{color}
\usepackage{amsmath}
\usepackage{multirow}
\newcommand{\greco}[1]{%
\begin{otherlanguage*}{greek}#1\end{otherlanguage*}}

\usepackage{array}
\usepackage{color}
\usepackage{colortbl}
\begin{document}

{\large
\title{A class of auxetic three-dimensional lattices}
}

\author{Luigi Cabras$^{a}$, Michele Brun$^{a,*}$}

\date{$^a${\em Dipartimento di Ingegneria Meccanica, Chimica e dei Materiali,} \\
{\em Universit\'a degli Studi di Cagliari, Piazza d'Armi, I-09123 Cagliari, Italy}
}


\maketitle

\vspace{10mm}\noindent
{\bf Abstract}

We propose a class of auxetic three-dimensional lattice structures. The elastic microstructure can be designed in order to have omni-directional Poisson's ratio arbitrarily close to the stability limit $-1$. The cubic behavior of the periodic system has been fully characterized; the minumum and maximum Poisson's ratio and the associated principal directions are given as a function of the microstructural parameters.

The initial microstructure is then modified into a body centered-cubic system that can achieve a Poisson's ratio lower than $-1$ and that can also behave as an isotropic three-dimensional auxetic structure.

\vspace{15mm}
$^*$ Corresponding author. Tel.: +39 070 6755411; fax:  +39 070 6755418. \\
$~${\em e-mail}: mbrun@unica.it; $~${\em web-page}: http://people.unica.it/brunmi/

\vspace*{5mm}
\noindent Keywords: Auxetic structure, microstructured medium, negative Poisson's ratio, elasticity, metamaterials.

\section{Introduction}


The Poisson's ratio $\nu$, the negative ratio between lateral and longitudinal deformations, is an indication of the capacity of a material to resist to different types of deformation.
For almost incompressible materials like rubber, most liquids and granular solids, external mechanical loads must spend much more work to change the volume rather than changing their shape. On the contrary re-entrant foams \cite{Lakes1987,FriisLakesPark1988} and some molecular structures \cite{YeganehWeidnerParise1992,Bau1998,HallColuciGalvaoKozlovZhangDantasBaughman2008} can easily change the volume homotetically, but they are hardly deformable in shape.
For a stable material the Poisson's ratio ranges between the `dilational' lower bound $-1$ and the `incompressible' upper bound $0.5$ in the linear isotropic case, while for anisotropic materials constitutive stability defines a domain in a properly defined $n-$dimensional space, as shown in \cite{GuoWheeler2006} and in \cite{Norris2006} for the cubic case. For a perfect `dilational' material a dilation is the only easy mode of deformation \cite{Mil2013a,Mil2013b}, which results in an isotropic material with $\nu=-1$.

Auxetic materials are materials with negative Poisson's ratio, expanding (contracting) laterally in one or more directions, when stretched (compressed) longitudinally. The term auxetic, from the Greek \greco{a>'uxhsic} (auxesis: increase, grow), was  firstly proposed by Evans \cite{Evans1991}.
Such non standard materials are increasingly used in sport applications (impact protector devices, better conformability for comfort and support, and enhanced energy absorption for lighter and/or thinner components) \cite{Sanami2014,GlazzardBreedon2014,WangHu2014}, in biomedicine (efficient design of dilators and stents, arterial prostheses and bandages) \cite{CaddockEvans1989,Gatt2014,MirAliSamiAnsari2014,AliBusfieldRehman2014,Martz2005}, fabric and textile \cite{AldersonRasburn2000,AldersonAlderson2005,ScarpaGiacominZhangPastorino2005} and military applications \cite{MaLiuLiu2011}.
The strong interest for engineering applications is related to enhanced properties, such as shear resistance \cite{ScarpaTomlin2000,JuSummers2011}, indentation resistance \cite{CoenenAlderson2011,AldersonFitzgeraldEvans2000}, synclastic curvature \cite{EvansAlderson2000}, crashworthiness \cite{ScarpaYatesCiffoPatsias2002}, pore-size tunability \cite{AldersonRasburn2000,AldersonRasburn2001} and sound absorption and vibration damping  \cite{ScarpaSmith2004,BaravelliaRuzzene2013,ScarpaCiffoYates2004,RuzzeneScarpa2003}.
Recently, negative Poisson's ratio materials have been also associated with metamaterials \cite{GattMizzi2014,Krodel2014,ShinUrzhumovLimKimSmith2014}; concerning such parallelism, it is important to correct a general misconception that identifies metamaterials to all manmade microstructure media; properly speaking metamaterials concern the dynamic behavior in the low-frequency regime \cite{CrasterGuenneau2013}.

While Love in his famous treatise  \cite{Love1944} already mentioned materials with negative Poisson's ratio, they started to become popular after the first artificial examples, such as the
re-entrant three-dimensional structure of
\cite{Almgren1985,Kolpakov1985}, the re-entrant polymeric foam of \cite{Lakes1987}, and the continuum models of \cite{Milton1992,MilCher1995}.
Extended reviews of existing models can be found in \cite{GreavesGreerLakesRouxel2011,ElipeLantada2012,MirAliSamiAnsari2014,Milton2015}.

A mechanism made of rotating rigid units can exhibit auxetic behaviour as proposed by \cite{Attard2012}; it is composed of rigid cuboids connected at their edges, which deform through relative rotation with respect to each other. 
The system exhibits negative values for all the six on-axis Poisson's ratios.
The use of three-dimensional models to predict or to explain auxetic behaviour is used in \cite{Evans2001,Evans2002,Evans2009} where a three-dimensional rotating and/or dilating tetrahedral model is applied to real crystalline auxetic materials such as $\alpha$-cristobalite and $\alpha$-quartz structures of both silica and germania.
Auxetic responses have been demonstrated possible in many crystals \cite{BauGalvao1993,Bau1998}, and some artificial three-dimensional auxetic materials have been proposed especially with the advent of the 3D-printer. A new material, named `Bucklicrystals', achieves a three-dimensional auxetic behavior  through the elastic buckling \cite{Bertoldi2013}. In \cite{Buckmann2014} it is described a dilational three-dimensional cubic auxetic material with an ultimate Poisson's ratio of $\nu=-1$, based on a two-dimensional chiral model recently published \cite{Mil2013a}; the three-dimensional model is studied numerically and a sample is fabricated with a 3D-printer.
The development of the modern 3D-printers has allowed the creation of metamaterials starting from theoretical models with unit cells in the micro-meter range: in \cite{Buckmann2012} the authors fabricated and characterized a truly three-dimensional crystalline mechanical metamaterial with unit cells in the micro-meter range and with adjustable Poisson's ratios, including negative values. They modified appropriately the DLW technology (Direct Laser Writing) to create three-dimensional nano- and micro-structures height only some tens of micro-meters, to get larger structures, and they apply this new approach to a model previously introduced in \cite{GibAsh1997}.

Recently we have proposed a class two-dimensional lattice models having the Poisson's ratio arbitrarily close to $-1$ \cite{CabrasBrun2014}; in this work we rediscovered a cubic model firstly given in \cite{Sigmund1994,Sigmund1995}, where an advanced topology optimization procedure was used in order to design the optimal structure. In \cite{CabrasBrun2014} we also generalized the cubic structure to an isotropic structures with omni-directional negative Poisson's ratio which remains arbitrarily close to the stability limit $-1$. In the model internal hinges avoid concentration of deformations that can easily lead to plastic deformations. In addition, the superposition of two systems rotating in opposite directions balances any internal couple canceling any chirality. Microstructured media with omni-directional negative Poisson's ratio have also been proposed in \cite{WangSigmundJensen2014} for plane structures.
Here, we extend such plane structures to a three-dimensional cubic lattice as in \cite{Sigmund1995} and we analyse in detail the effective behavior considering elastic elements and additional internal springs leading to overall stability. The initial system has a simple cubic unit cell with strong directional dependance of the effective constitutive properties. A modified body-centered cubic microstructure is designed in order to have omni-directional negative Poisson's ratio.

The paper is organized as follows. In Section \ref{Cubic Section} we show the simple cubic auxetic lattice and we analyse the kinematics of the unimode structure. The effective properties are determined in Section \ref{Section2.1} and \ref{Section2.2}, where we consider first  the behavior along the principal directions of the cubic system and then we search for the extremal values of the Young's modulus and the Poisson's ratio. The structure shows strong directional dependance and a modified body centered cubic structure, stiffened with respect to shear deformations, is presented in Section \ref{Sect3}. Such modified structure can be tailored to give isotropic behavior or Poisson's ratio less than $-1$ along particular directions. Final considerations conclude the paper.

\section{The auxetic lattice}
\label{Cubic Section}

\begin{figure}[!htcb]
\centerline{
	\begin{tabular}{c}
         	 	\begin{tabular}{c@{\hspace{0.5pc}}c@{\hspace{0.5pc}}c}
             		\includegraphics[width=0.5\columnwidth]{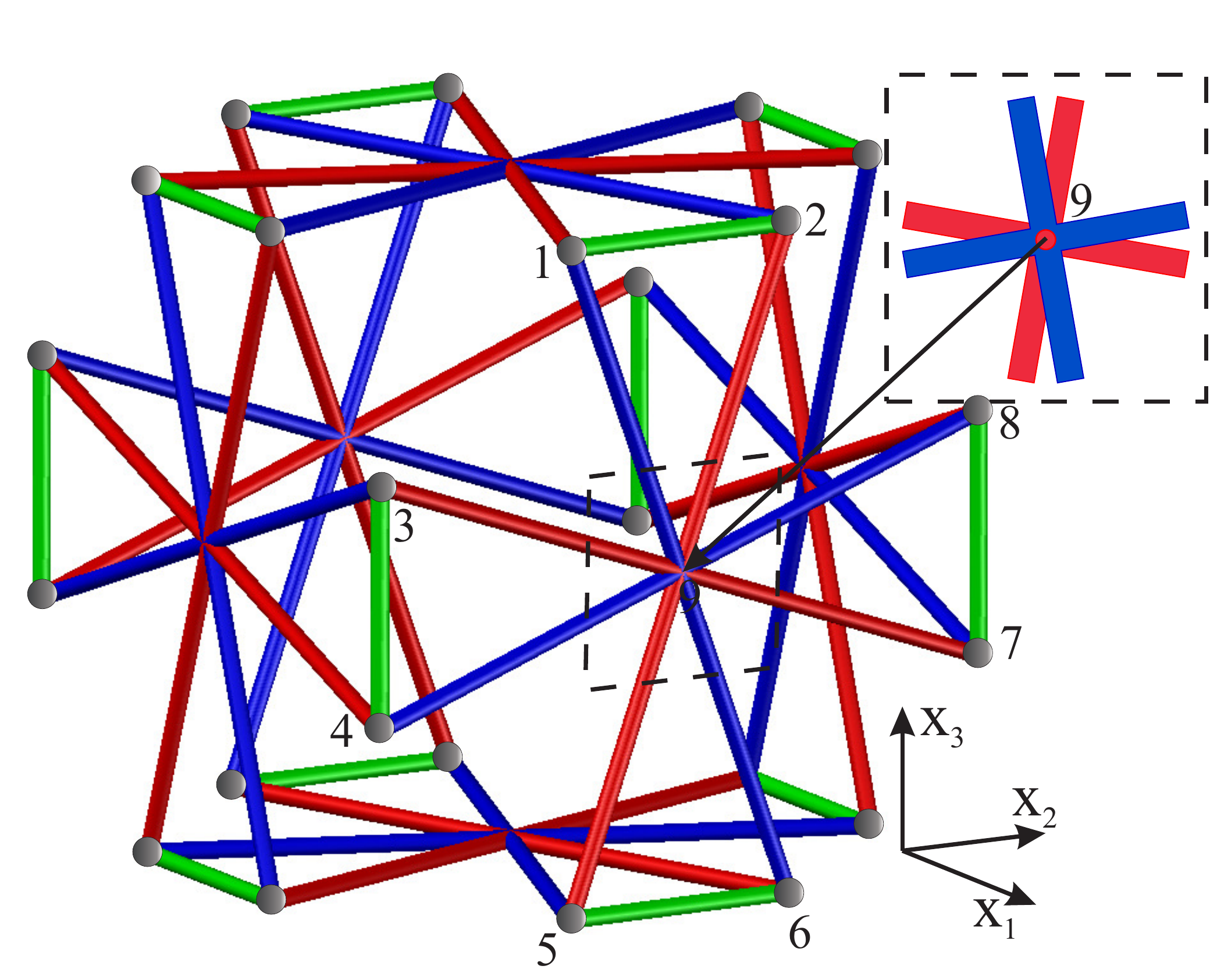} &
             		\includegraphics[width=0.4\columnwidth]{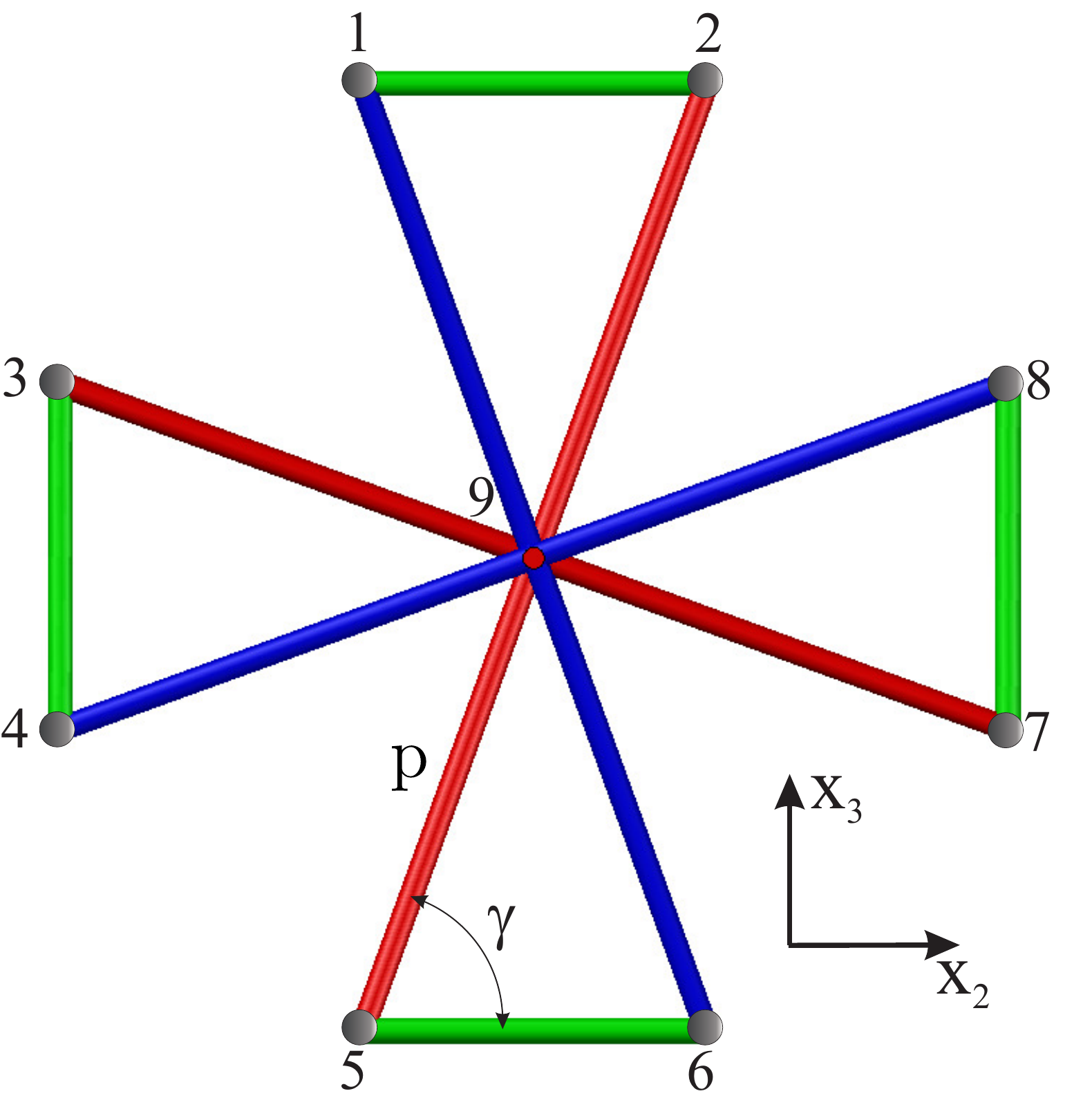}\\
         			(a) & (b)
         		\end{tabular} \\
                  \end{tabular}
       }
\caption{(a) Three-dimensional cell of the cubic lattice. (b) Particular of a face of the typical cell of the cubic lattice. The numbers identify the node in part (a) and (b).}
\label{fig2_1_1}
\end{figure}

The typical cell of the lattice is presented in Fig. \ref{fig2_1_1}a. It is composed of slender cross-shaped elements, represented in red and blue, joined by a hinge at the central point, as in Fig. \ref{fig2_1_1}b. Additional trusses, represented in green, connect the cross-shaped elements at the end points.  Introducing the terminology of crystal structure, the unit cell is a cubic P (cP) Bravais or space lattice \cite{Kittel1996} with cubic symmetry.
In Fig. \ref{fig2_1_1}b we indicate the geometrical parameters characterizing the structure: $p$ is the length of the arms of the cross-shaped elements and $\gamma$ their inclination with respect to the truss elements. The longitudinal truss elements have cross section $A_t$, Young's modulus $E_t$ and longitudinal stiffness $k_t=E_tA_t/(2p\cos\gamma)$. The arms of the cross-shaped elements are considered as Euler beams with cross section $A_c$, second-moment of inertia $J_c$ and Young's modulus $E_c $.

\begin{figure}[!htcb]
\centerline{
	\begin{tabular}{c}
         	 	\begin{tabular}{c@{\hspace{0.5pc}}c@{\hspace{0.5pc}}c}
             		\includegraphics[width=0.45\columnwidth]{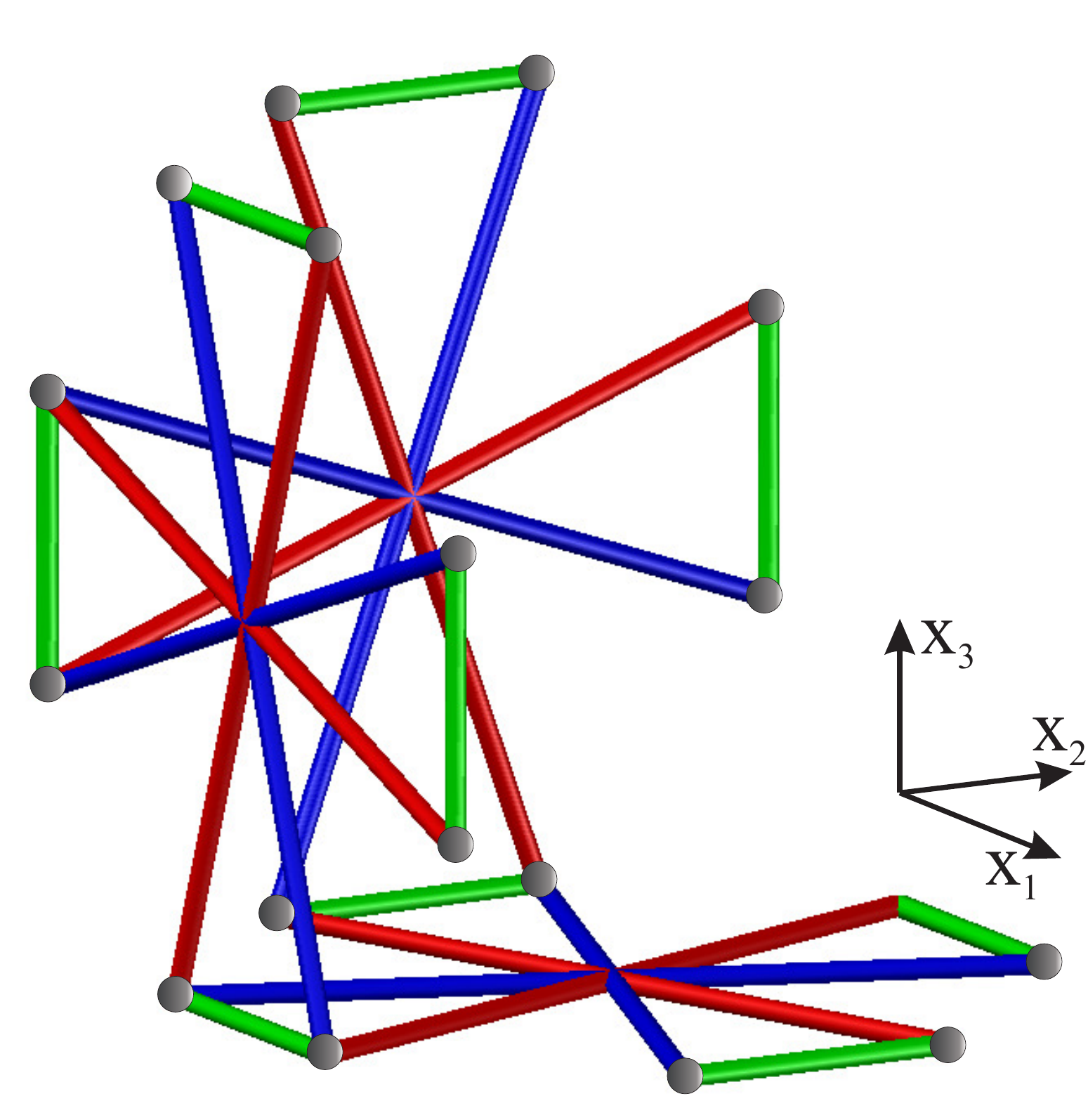} &
             		\includegraphics[width=0.45\columnwidth]{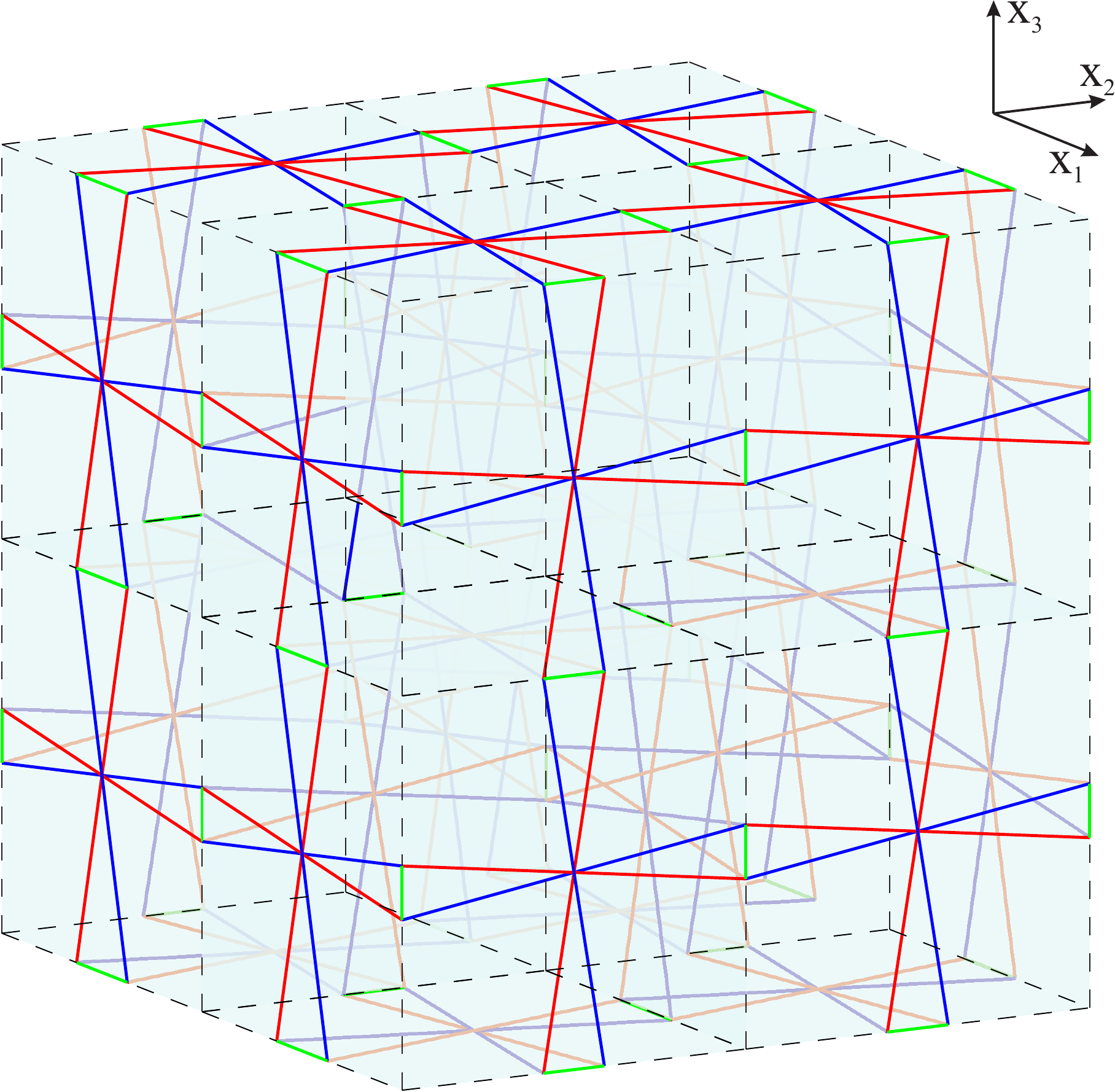}\\
         			(a) & (b)
         		\end{tabular} \\
                  \end{tabular}
       }
\caption{(a) Unit cell implemented in Comsol Multiphysics\textsuperscript{\textregistered}. (b) Three-dimensional cubic lattice.}
\label{fig2}
\end{figure}

We tesselate the space starting from the unit cell in Fig. \ref{fig2}a. The corresponding periodic structure, given in Fig.  \ref{fig2}b, has a Bravais periodic lattice \cite{Kittel1996} consisting of points
\begin{equation}
{\bf R}= n_1{\bf t}_1+n_2{\bf t}_2+n_3{\bf t}_3,
\label{eqn001}
\end{equation}
where $n_1, n_2, n_3$ are integers and
\begin{equation}
{\bf t}_1=
\begin{pmatrix}
2p\sin\gamma\\
0\\
0\\
\end{pmatrix}, \qquad
{\bf t}_2=
\begin{pmatrix}
0\\
 2p\sin\gamma\\
0\\
\end{pmatrix}, \qquad
{\bf t}_3=
\begin{pmatrix}
0\\
0\\
 2p\sin\gamma\\
\end{pmatrix},
\label{eqn002}
\end{equation}
are the primitive vectors spanning the lattice.

In the absence of trusses and for rigid cross-shaped elements we refer to \cite{CabrasBrun2014} Section 3(a) and to \cite{PatAna2007} for the description of the kinematics of each couple of elements joined at the central points. In particular, the only possible mode of deformation is homothetic expansion or contraction.
Following the systematic analysis for finite deformation given in \cite{MilCher1995,Mil2002,Mil2013a,Mil2013b} we show that the lattice is a unimode material. Let
\begin{equation}
{\bf T}=\left[\begin{array}{ccc}
{\bf t}_1 &  {\bf t}_2 &  {\bf t}_3
\end{array}
\right] = 2p\sin\gamma \, {\bf I}
\label{eqn003}
\end{equation}
be the `lattice matrix', where ${\bf I}$ is the identity matrix.
During the deformation the primitive vectors undergo an affine transformation and the matrix ${\bf T}$ describes a motion starting at $t=t_0$, with $\gamma(t_0)=\gamma_0$.
At time $t$ the deformation gradient is  given by
\begin{equation}
{\bf F}(t,t_0)=[{\bf T}(t)][{\bf T}(t_0)]^{-1}.
\label{eqn102}
\end{equation}
Then, the corresponding Cauchy-Green deformation tensor is the path
\begin{equation}
{\bf C}(t,t_0)=[{\bf T}(t_0)]^{-T}[{\bf T}(t)]^T[{\bf T}(t)][{\bf T}(t_0)]^{-1}=\left(\frac{\sin\gamma}{\sin\gamma_0}\right)^{2} {\bf I}.
\label{eqn004}
\end{equation}
As for the planar square lattice in \cite{CabrasBrun2014} the only possible path ${\bf C}(t,t_0)$ lie on a one-dimensional curve and we can conclude that the lattice structure is unimode.

\subsection{Effective properties}
\label{Section2.1}

The lattice structure has cubic symmetry and its macroscopic behavior is described by three independent elastic moduli. The elements of the lattice are sufficiently slender, so that classical structural theories can be conveniently applied to analyse the effective response.
The structure has been studied numerically with the finite element code, Comsol Multiphysics\textsuperscript{\textregistered}. We have considered the following material and geometrical parameters:  steel with Young's modulus $E_c=200000$ MPa, cross-shaped elements with length of the arms $p=10$ mm and circular cross section having radius $r=0,25$ mm, so that the area is $A_c=0,196$ mm$^2$.
The truss elements have longitudinal stiffness $k_L$ and we introduce the non-dimensional stiffness ratio parameter $\alpha=k_Lp/(E_cA_c)$, between the longitudinal stiffness of the trusses and of the arms of the cross-shaped elements.

\begin{figure}[htbp]
\centering
\vspace*{1mm} \rotatebox{0}{\resizebox{!}{9cm}{%
\includegraphics[scale=1]{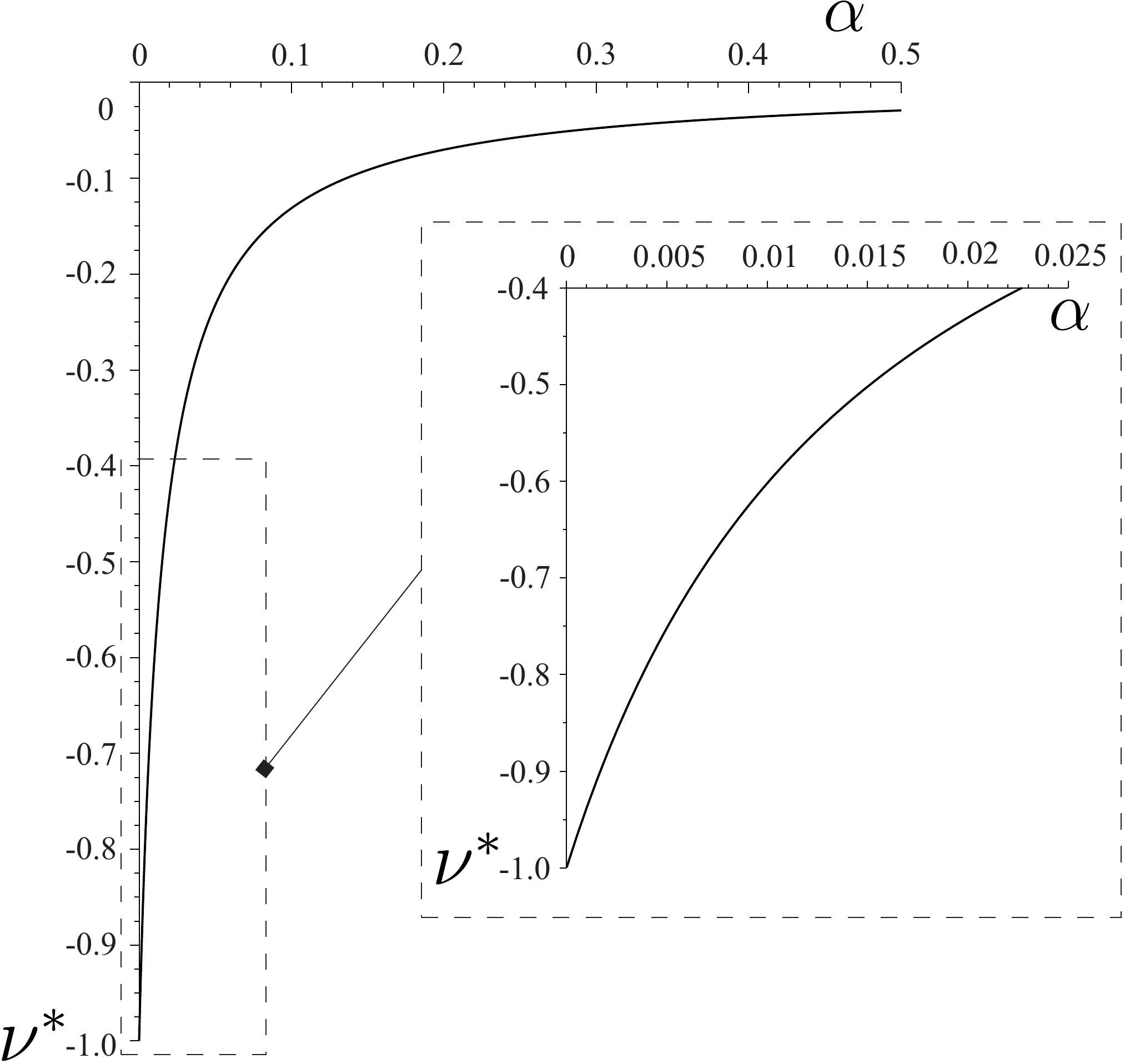}}}
 \caption{Poisson's ratio $\nu^*$ as a function of the non-dimensional stiffness ratio $\alpha=k_Lp/(E_cA_c)$.}
\label{fig2_1_9}
\end{figure}

The single unit cell has been implemented, subjected to periodic boundary conditions. In particular, displacements ${\bf u}=(u_1,u_2,u_3)$ along the principal directions ${\bf t}_1$, ${\bf t}_2$, ${\bf t}_3$ of the lattice have been applied on the boundary and the corresponding reaction forces on the lattice have been obtained. Displacements  and reaction forces are associated to macroscopic strains $\bar {\bf \varepsilon}$ and macroscopic stresses $\bar \sigma$, respectively.
The effective Poisson's ratio
\begin{equation}
\nu^*=\frac{\bar\sigma_{22}\bar\varepsilon_{11}-\bar\sigma_{11}\bar\varepsilon_{22}}
{(\bar\sigma_{11}-\bar\sigma_{22})\bar\varepsilon_{33}+(\bar\sigma_{11}+\bar\sigma_{22})(\bar\varepsilon_{11}-\bar\varepsilon_{22})},
\label{eqn200}
\end{equation}
is given in Fig. \ref{fig2_1_9} as a function of the stiffness ratio $\alpha$. \footnote{Note that equivalent definitions of $\nu^*$, $E^*$ and $\mu^*$, are given by permutation of the indexes $1,2,3$ in Eqs. (\ref{eqn200}-\ref{eqn202})}
It is shown that the lattice has auxetic behavior in the whole range of $\alpha$. In the limit when $\alpha\to 0$ the Poisson's ratio $\nu^*$ is arbitrarily close to the stability limit $-1$.

\begin{figure}[htbp]
\centerline{
	\begin{tabular}{c}
         	 	\begin{tabular}{c@{\hspace{0.5pc}}c@{\hspace{0.5pc}}c}
             		\includegraphics[width=0.45\columnwidth]{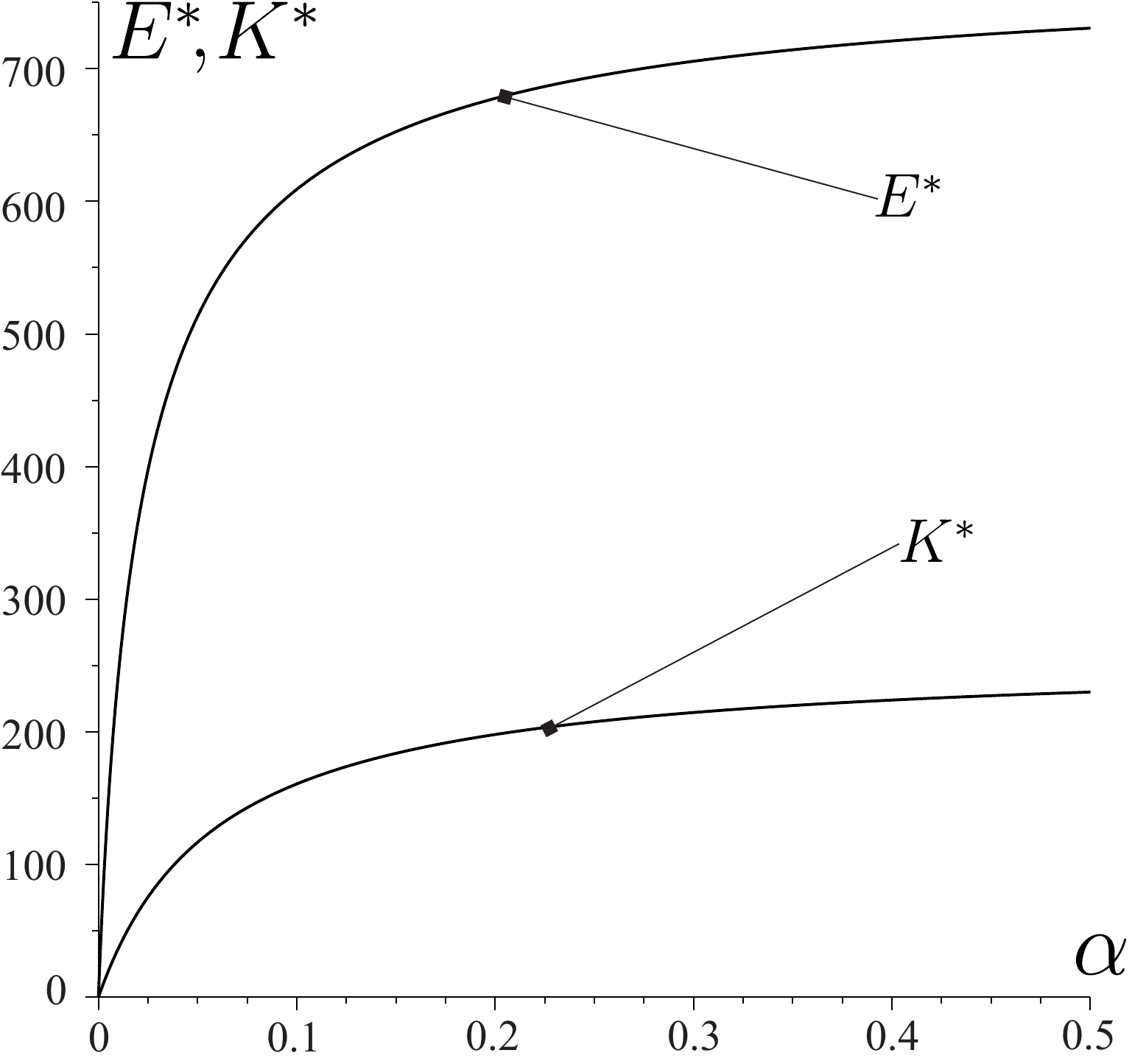} &
             		\includegraphics[width=0.45\columnwidth]{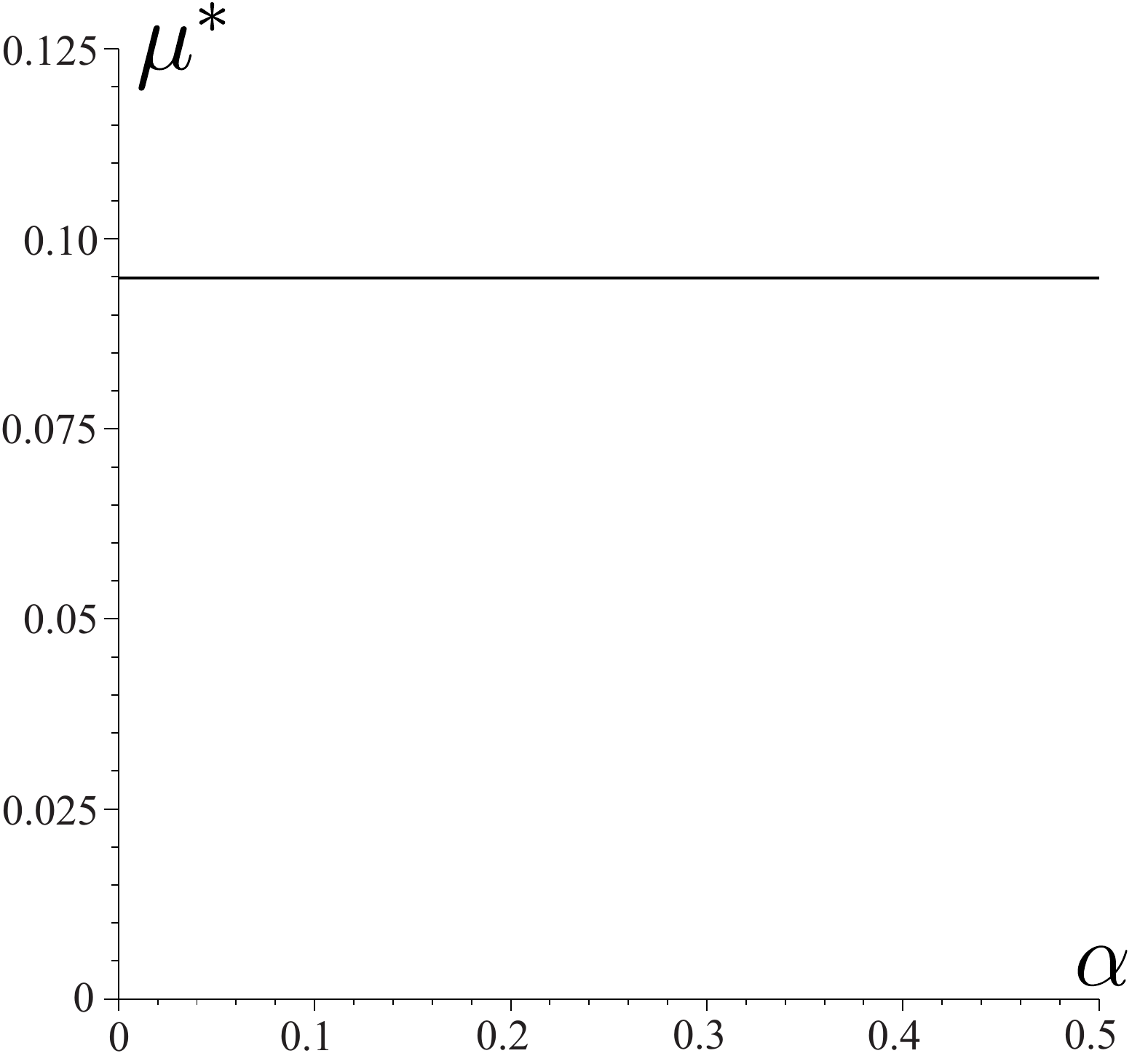}\\
         			(a) & (b)
         		\end{tabular} \\
                  \end{tabular}
       }
\caption{(a) Effective Young's modulus $E^*$ and effective bulk modulus $K^*$. (b) Effective shear modulus $\mu^*$. Elastic moduli are given as a function of the non-dimensional stiffness ratio $\alpha$.}
\label{fig2_1_6}
\end{figure}

The Young's modulus
\begin{equation}
E^*=\frac{(\bar\sigma_{11}-\bar\sigma_{22})[\bar\sigma_{11}(\bar\varepsilon_{11}-2\bar\varepsilon_{22}+\bar\varepsilon_{33})-\bar\sigma_{22}(-2\bar\varepsilon_{11}+\bar\varepsilon_{22}+\bar\varepsilon_{33})]}
{(\bar\sigma_{11}-\bar\sigma_{22})\bar\varepsilon_{33}+(\bar\sigma_{11}+\bar\sigma_{22})(\bar\varepsilon_{11}-\bar\varepsilon_{22})}
\label{eqn201}
\end{equation}
and the bulk and the shear moduli
\begin{equation}
K^*=\frac{\mbox{Tr} (\bar{\bf \sigma}) }{3\,\mbox{Tr} (\bar{\bf \varepsilon})}, \qquad
\mu^*=\frac{\bar\sigma_{12}}{2\bar\varepsilon_{12}},
\label{eqn202}
\end{equation}
are shown in Fig. \ref{fig2_1_6}

We note that in the limit $\alpha\to 0$ the Young's and the bulk moduli tend to zero while the shear modulus remains finite (and positive).
This is clearly fully consistent with the linear theory of elasticity for cubic materials and also observed asymptotically for the plane model in \cite{CabrasBrun2014}.
In particular, the elasticity tensor in case of cubic symmetry can be expressed as
\begin{equation}
\mathbb{C}=3K^*\mathbb{J}+2{\hat{\mu}}^*{\hat{\mathbb{K}}}+2\mu^*\mathbb{K},
\label{eqn203}
\end{equation}
where $\hat\mu^*$ is a second shear modulus for shear inclined at $45^\circ$ with respect to two principal directions of the lattice \cite{Thomas1965,Walpole1984}.
The fourth-order projection tensors in \eq{eqn203} are $\mathbb{J}=(1/3) {\bf I}\otimes{\bf I}$, with ${\bf I}$ second-order identity tensor, ${\hat{\mathbb{K}}}=\mathbb{T}-\mathbb{J}$ and $\mathbb{K}=\mathbb{I}-\mathbb{T}$. The tensors $\mathbb{I}$ and $\mathbb{T}$ have components
\begin{equation}
I_{ijkl}=\frac{1}{2}(\delta_{ik}\delta_{jl}+\delta_{il}\delta_{jk}), \quad
T_{ijkl}=\delta_{ij}\delta_{jk}\delta_{kl},  \qquad (i,j,k,l \mbox{ not summed}).
\label{eqn204}
\end{equation}
The Poisson's ratio is then given by
\begin{equation}
\nu^*=\frac{3K^*-2\hat\mu^*}{2(3K^*+\hat\mu^*)},
\label{eqn205}
\end{equation}
confirming that $\nu^*\to -1$ when $K^*\to 0$ .

From the results in \fig{fig2_1_6}b we also note that the shear modulus $\mu^*$ is independent of the ratio $\alpha$, this is due to the fact that during shear deformation the longitudinal truss elements are not deformed and the effective stiffness of the system is associated to longitudinal deformations of the arms of the cross-shaped elements, a deformation mechanism already observed in the plane model \cite{CabrasBrun2014}.

\subsection{Directional dependance of effective properties}
\label{Section2.2}

In the previous Section the effective properties have been shown in the principal system of the lattice. Here we detail the directional dependance of the elastic moduli, in particular we focus on the Poisson's ratio and the Young's modulus. It is convenient to consider the compliance tensor $\mathbb{S}={\mathbb{C}}^{-1}$, which, for cubic material, in addition to the major and minor symmetries $S_{ijkl}=S_{klij}$ and $S_{ijkl}=S_{jikl}=S_{ijlk}$, has
\begin{equation}
S_{1111}=S_{2222}=S_{3333}, \quad
S_{1122}=S_{2233}=S_{3311}, \quad
S_{1212}=S_{2323}=S_{3131},
\label{equ301}
\end{equation}
with the constrained of positive definiteness \cite{Ting2005}
\begin{equation}
S_{1111}-S_{1122}>0,\quad S_{1111}+2S_{1122}>0,\quad S_{1212}>0.
\label{equ302}
\end{equation}
In a rotated frame of reference, where the rotation is described by the proper orthogonal tensor ${\bf Q}^T$, the new components $S_{ijkl}'$ of the compliance tensor are
\begin{equation}
S_{ijkl}'=Q_{ip}Q_{jq}Q_{kr}Q_{ls}S_{pqrs}.
\label{equ303}
\end{equation}
The elastic compliance $S'_{1111}$ and $S'_{1122}$ are:
\begin{equation}
S'_{1111}= n_i n_j n_k n_l S_{ijkl} =S_{1111}(n^4_1+n^4_2+n^4_3)+(2S_{1122}+S_{1212})(n^2_1n^2_2+n^2_1n^2_3+n^2_2n^2_3),\\
\label{equ304}
\end{equation}
\begin{eqnarray}
\nonumber
S'_{1122}&= n_i n_j m_k m_l S_{ijkl} =S_{1111}(n^2_1m^2_1+n^2_2m^2_2+n^2_3m^2_3)+S_{1122}(m^2_1n^2_2+m^2_2n^2_1+m^2_1n^2_3+\\
&m^2_3n^2_1++m^2_2n^2_3+m^2_3n^2_2)+S_{1212}(n_1n_2m_1m_2+n_1n_3m_1m_3+n_2n_3m_2m_3),
\label{equ305}
\end{eqnarray}
where ${\bf n}$ and ${\bf m}$ are orthogonal unit vectors. 

The dependance of the Young's modulus on the direction, $E^*=E^*(\textbf{n})$, can be obtained from the relation
\begin{equation}
\frac{1}{E^*(\textbf{n})}=S'_{1111}, 
\label{equ306}
\end{equation}
which, making use of Eq. (\ref{equ304}) and the identity
\begin{equation}
n^4_1+n^4_2+n^4_3=1-2(n^2_1n^2_2+n^2_1n^2_3+n^2_2n^2_3),
\label{equ307}
\end{equation}
gives
\begin{equation}
E^*(\textbf{n})=\left[S_{1111}-(2S_{1111}-2S_{1122}-S_{1212})(n^2_1n^2_2+n^2_1n^2_3+n^2_2n^2_3)\right]^{-1}.
\label{equ308}
\end{equation}

The directional dependance of the Poisson's ratio $\nu^*(\textbf{n},\textbf{m})$ is obtained from the relation
\begin{equation}
\nu^*(\textbf{n},\textbf{m}) =-\frac{S'_{1122}}{S'_{1111}}.
\label{equ309}
\end{equation}
We rewrite $S'_{1122}$ in Eq. (\ref{equ305}) making use of the following identities:
\begin{align}
1-(n^2_1m^2_1+n^2_2m^2_2+n^2_3m^2_3)&=(m^2_1n^2_2+m^2_2n^2_1+m^2_1n^2_3+m^2_3n^2_1+m^2_2n^2_3+m^2_3n^2_2),\notag\\
(n^2_1m^2_1+n^2_2m^2_2+n^2_3m^2_3)&=-2(n_1n_2m_1m_2+n_1n_3m_1m_3+n_2n_3m_2m_3),
\label{equ310}
\end{align}
so that we have the simplified expression
\begin{equation}
S'_{1122}=S_{1122}+(S_{1111}-S_{1122}-S_{1212}/2)(n^2_1 m^2_1+n^2_2m^2_2+n^2_3m^2_3).
\label{equ311}
\end{equation}
Therefore, the directional dependance of the Poisson's ratio $\nu^*(\textbf{n},\textbf{m})$ is
\begin{equation}
\nu^*(\textbf{n},\textbf{m}) =-\frac{S'_{1122}}{S'_{1111}}=- \frac{S_{1122}+(S_{1111}-S_{1122}-S_{1212}/2)(n^2_1 m^2_1+n^2_2m^2_2+n^2_3m^2_3)}{S_{1111}-2(S_{1111}-S_{1122}-S_{1212}/2)(n^2_1 n^2_2+n^2_2n^2_3+n^2_3n^2_1 )}.
\label{equ312}
\end{equation}

\subsubsection{Effective Young's modulus}

\begin{figure}[htbp]
\centerline{
	\begin{tabular}{ccc}
             		\includegraphics[width=0.45\columnwidth]{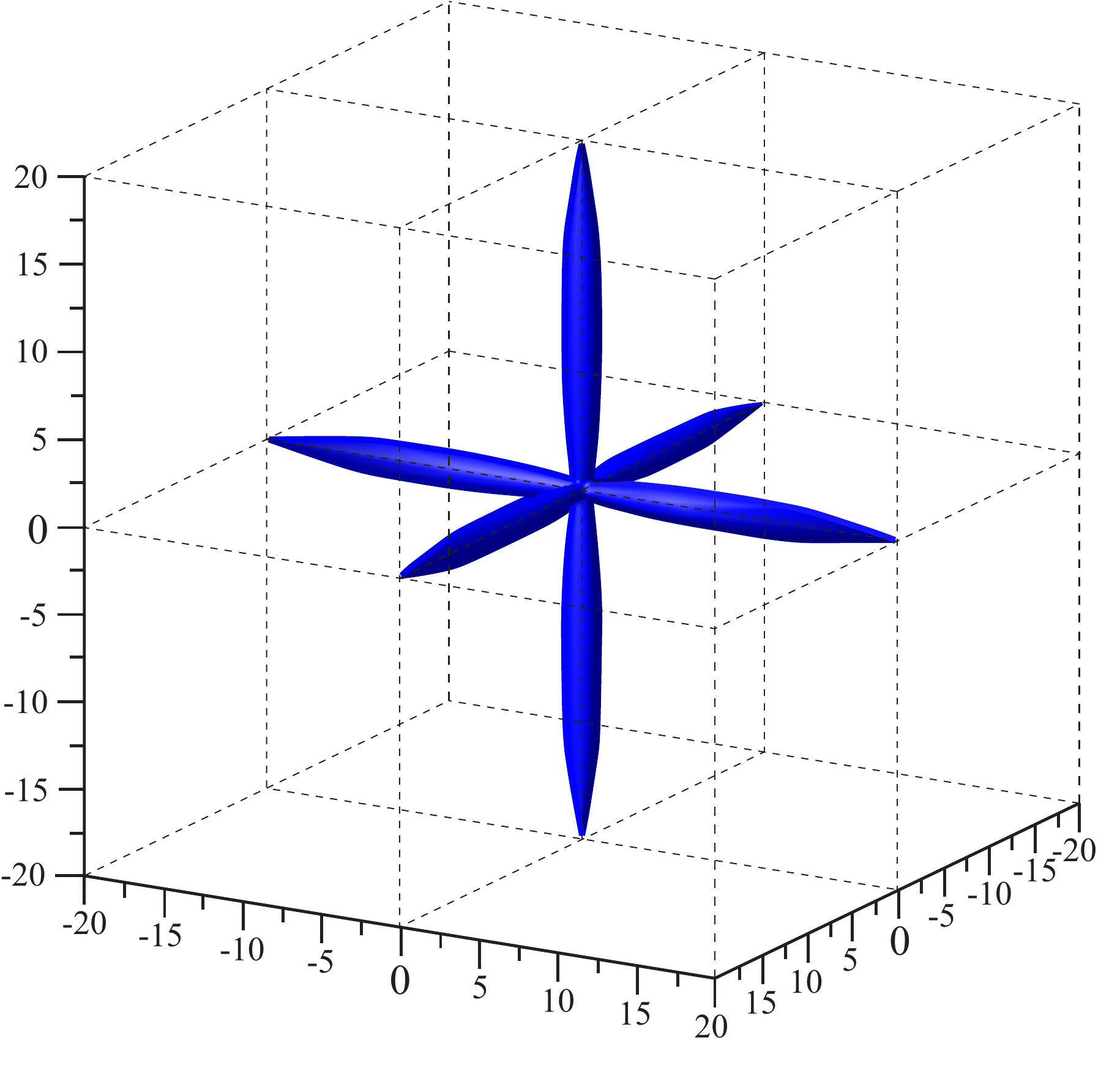} & &
             		\includegraphics[width=0.4\columnwidth]{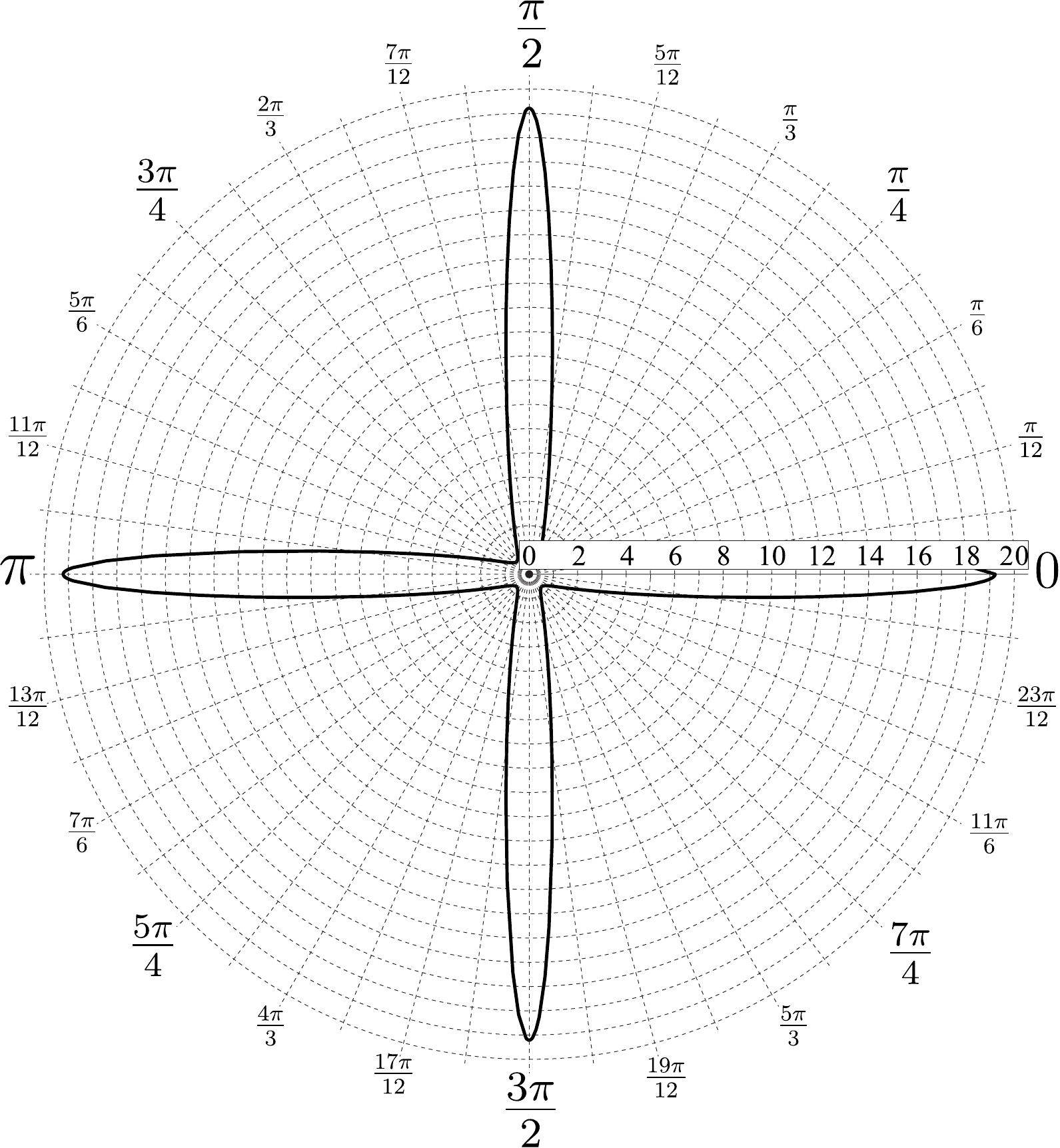}\\
         			(a) & & (b)
                  \end{tabular}
       }
\caption{(a) Polar representation of the effective Young's modulus $E^*({\bf n})$. (b) Polar diagram of the effective Young's modulus $E^*$ in the $[1a0]$ plane. Results are given for $\alpha=0.000515$.}
\label{fig2_2_7}
\end{figure}

The Young's modulus $E^*$ is shown in the polar plots in Fig. \ref{fig2_2_7} as a function of the direction ${\bf n}$. For low stiffness ratio $\alpha$, which is equal to $0.000515$ in Fig. \ref{fig2_2_7}, the structure has strong anisotropy highlighted by the strong variation of $E^*$, namely the ratio $E^*_{max}/E^*_{min}\gg 1$. Within the cubic behavior, the anisotropy can be quantified by the dimensionless Zener anisotropy factor \cite{Cazzani2003,Zener1948}
\begin{equation}
\beta_{cub}=\frac{2(S_{1111}-S_{1122})}{4S_{1212}},
\label{equ313}
\end{equation}
which is equal to $1$ for isotropic materials, while $\beta_{cub}=0.325\cdot 10^{-3}$ in Fig. \ref{fig2_2_7}.

\begin{figure}[!htcb]
\centerline{
         \begin{tabular}{c c}
                \includegraphics[width=6.5 cm]{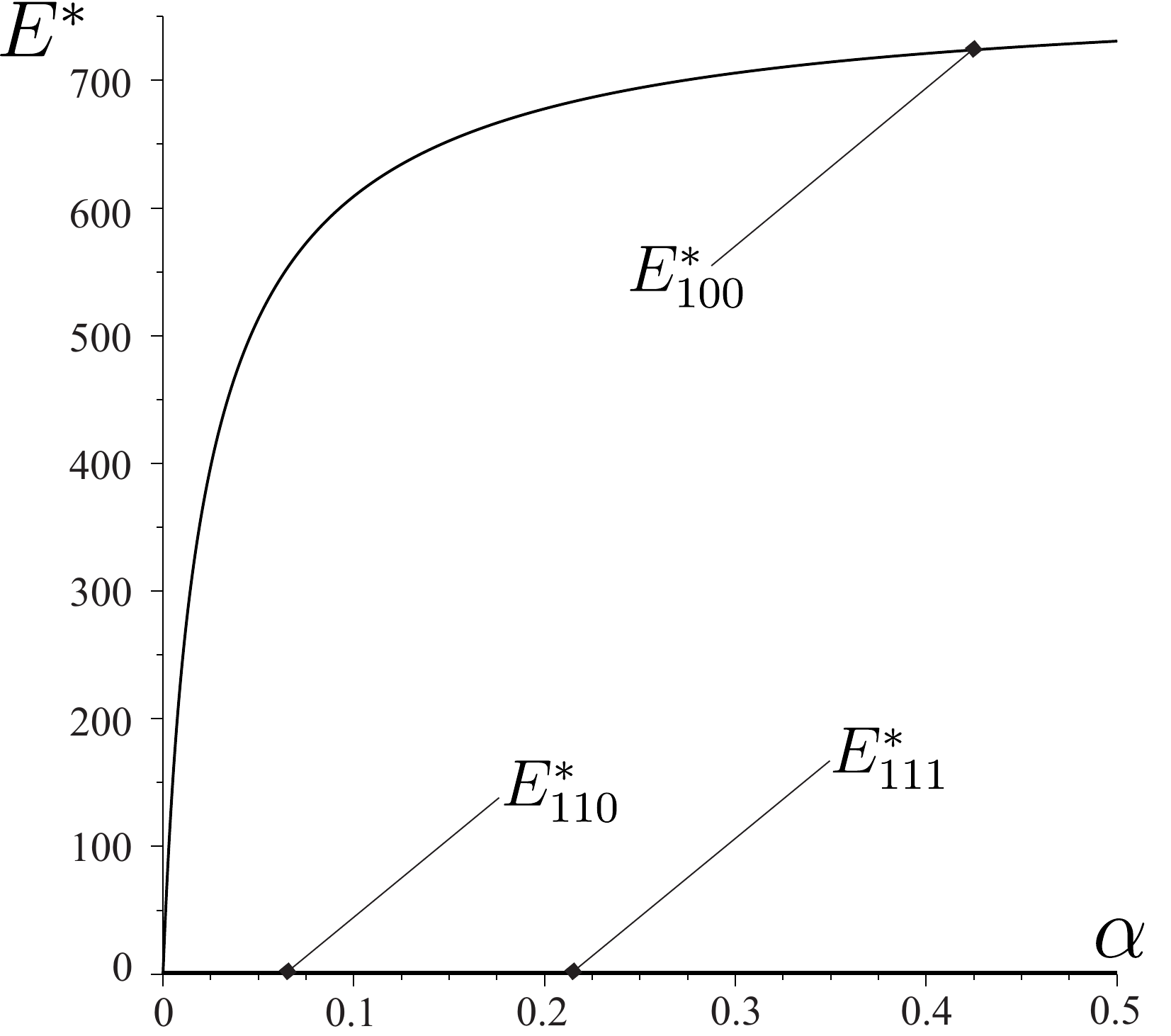} &
                \includegraphics[width=6.5cm]{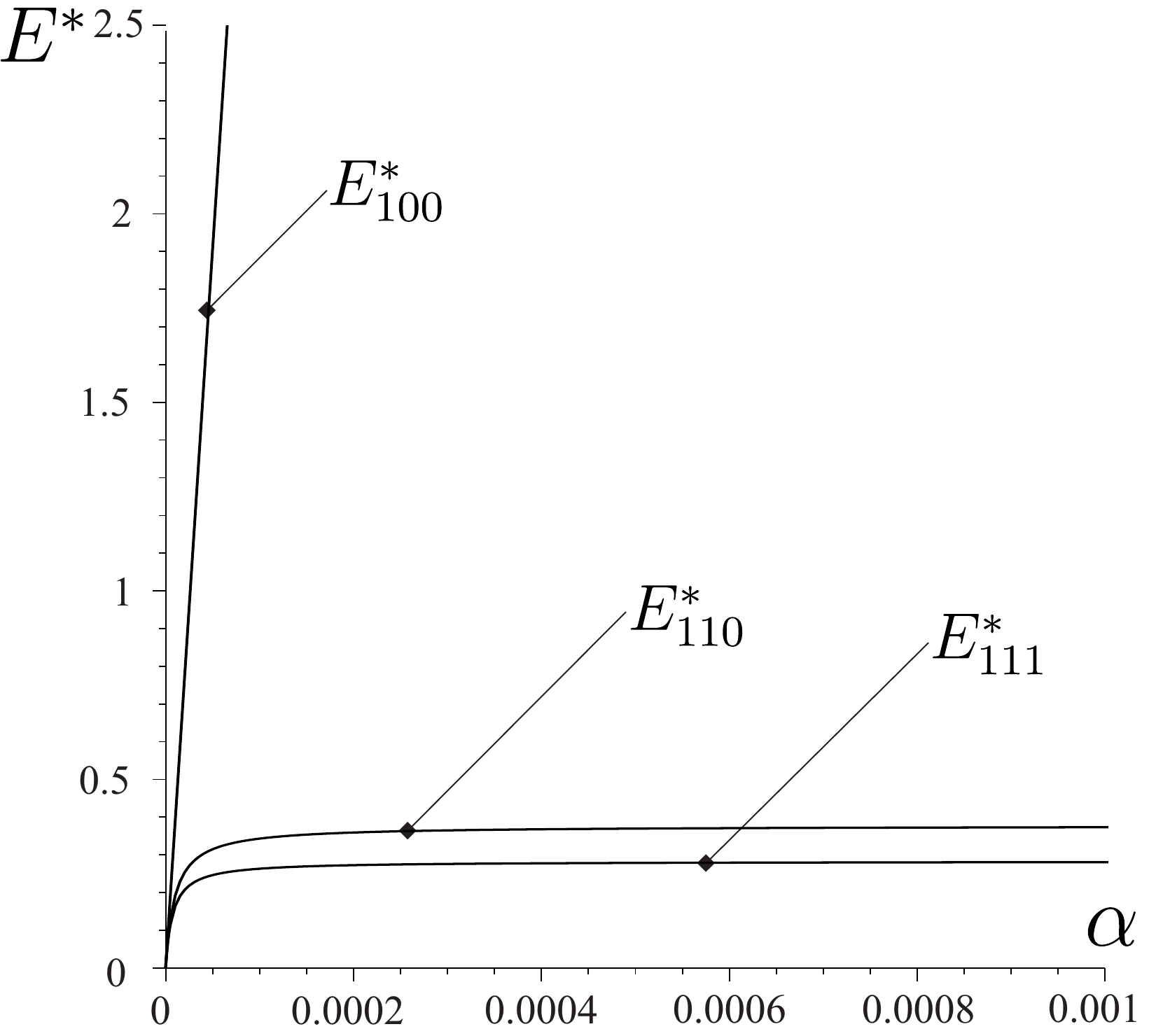} \\
         			(a)&(b)
			\\
         \end{tabular}
}
\caption{Effective Young's modulus $E^*$, along $[100]$, $[110]$ and $[111]$ directions, as a function of the non-dimensional stiffness ratio $\alpha$. (a) Effective Young's modulus for $\alpha\in(0,0.5)$. (b) Effective Young's modulus for $\alpha\in(0,0.001)$.}
\label{fig2_2_8}
\end{figure}

The extreme values of the Young's modulus are $E^*_{100}$, $E^*_{110}$ and $E^*_{111}$, where the subscripts indicate the crystallographic directions. They are represented in Fig. \ref{fig2_2_8} as a function of the stiffness ratio $\alpha$.
In particular, the global extrema are
\begin{eqnarray}
\nonumber
\beta_{cub}<1:\,\left\{
\begin{array}{l}
\displaystyle{
E^*_{max}=E^*_{100}=\frac{1}{S_{1111}}},\\[3. mm]
\displaystyle{E^*_{min}=E^*_{111}=\frac{3}{S_{1111}+2S_{1122}+4S_{1212}}},
\end{array}
\right. \\
\beta_{cub}>1:\,\left\{
\begin{array}{l}
\displaystyle{E^*_{max}=E^*_{111}=\frac{3}{S_{1111}+2S_{1122}+4S_{1212}}},\\[3.5 mm]
\displaystyle{E^*_{min}=E^*_{100}=\frac{1}{S_{1111}}},
\end{array}
\right.
\label{equ314}
\end{eqnarray}
as confirmed by the results in Fig. \ref{fig2_2_8}, where $\beta_{cub}<0.325\cdot 10^{-3}$.

\subsubsection{Effective Poisson's ratio}

While for linear isotropic stable materials Poisson's ratio is strictly bounded between $-1$ and $1/2$, such bounds do not exist for anisotropic solids, even for those `closest' to isotropy such as the cubic material. In \cite{Ting2005} it is demonstrated that arbitrarily large positive and negative values of Poisson's ratio could occur in solids with cubic material symmetry along specific directions. The key requirement is that $E_{111}^*$ is very large relative to other directions, and, as a consequence, the Poisson's ratio for stretch close to but not coincident with the $[111]$ direction can be large, positive or negative. \cite{Ting2005} also replaced the conventional belief  that the extreme values of $\nu^*$ are associated with stretch along the face diagonal $[110]$ direction (see, for example \cite{Bau1998}).

\begin{figure}[htbp]
\centering
\vspace*{1mm} \rotatebox{0}{\resizebox{!}{8.5cm}{%
\includegraphics[scale=1]{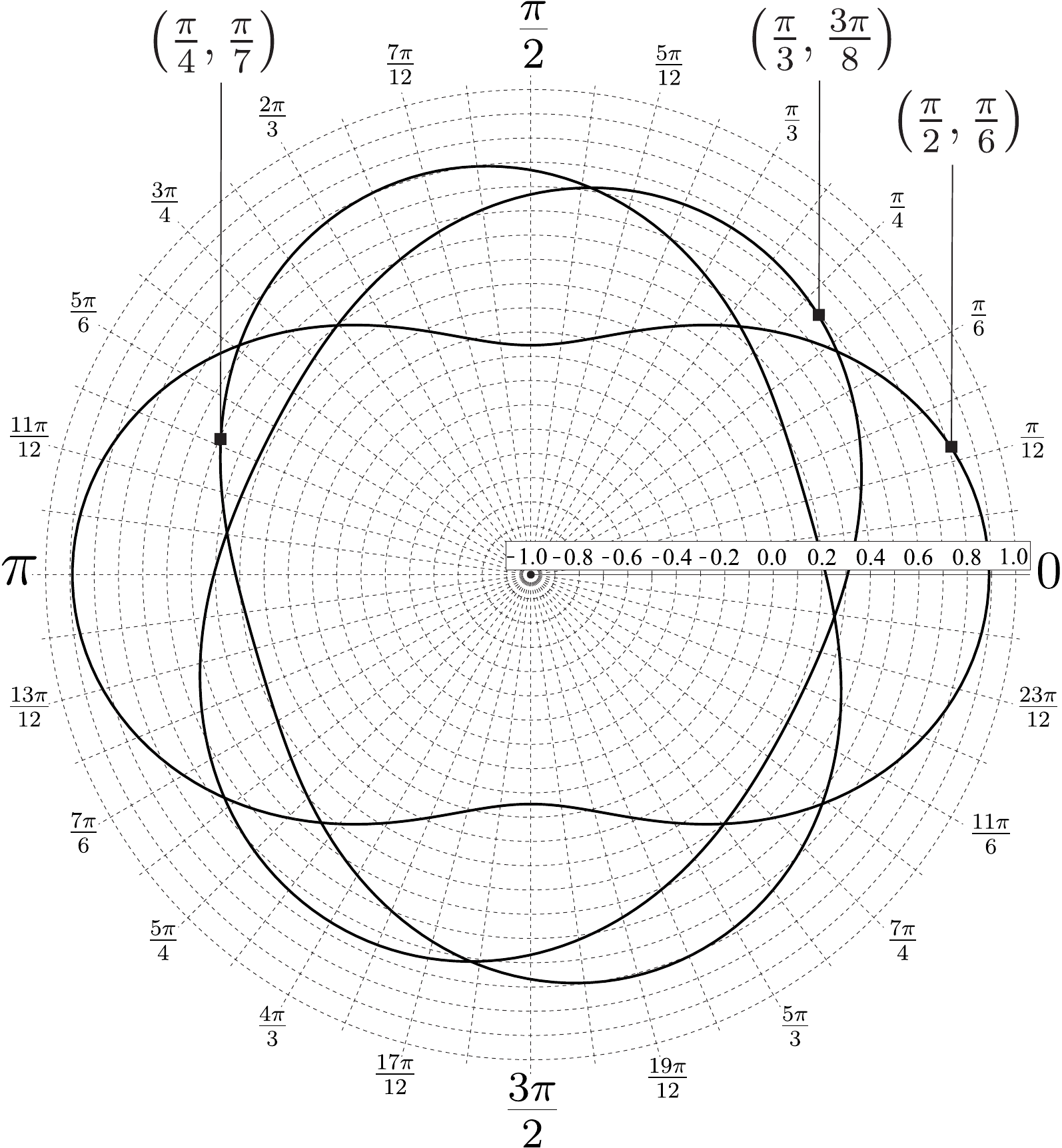}}}
 \caption{Polar diagrams of the Poisson's ratio $\nu^*$ as a function of the transverse direction \textbf{m}. Results are given for three different values of \textbf{n} described by the polar angle $\beta$ and the azimuthal angle $\phi$, $(\beta,\phi)=(\pi/2,\pi/6)$, $(\pi/3,3\pi/8)$, $(\pi/4,\pi/7)$ and for stiffness ratio $\alpha=0.000515$.}
\label{fig2_1_10}
\end{figure}

The Poisson's ratio $\nu^*(\textbf{n},\textbf{m})$ depends both on the direction of the applied uniaxial load \textbf{n} and on the transverse direction \textbf{m}, with $\textbf{m}\cdot\textbf{n}=0$.
Indeed, to find the extreme values, it is necessary to consider general \textbf{n} and \textbf{m}; for the purpose of illustration the dependance of $\nu^*$ on the direction $\textbf{m}$ for three different $\textbf{n}$ is shown in Fig. \ref{fig2_1_10}, where we have introduced the polar angle $\beta=\arccos(n_3)$ and the azimuthal angle $\phi=\arctan(n_2/n_1)$.

For the analysis of the maximum and minimum Poisson's ratio we refer to \cite{Ting2005,Norris2006}. In particular, if $\textbf{n}$ is supposed to be given, \textbf{m} depends on just one parameter and so the \textbf{m} that gives the extremes of $\nu^*(\textbf{n},\textbf{m})$ is:
\begin{equation}
\textbf{m} = \frac{\cos\theta}{\sqrt{1-n^2_3}}
\begin{bmatrix}
n_2\\
-n_1\\
0\\
\end{bmatrix}+ \frac{\sin\theta}{\sqrt{1-n^2_3}}
\begin{bmatrix}
-n_1n_3\\
-n_2n_3\\
1-n^2_3\\
\end{bmatrix}, \quad n^2_3\neq1,
\label{equ315}
\end{equation}
where $\theta$ is arbitrary. Inserting \eq{equ315} into \eq{equ312} leads to the condition of stationarity of $\nu^*(\textbf{n},\textbf{m})$
\begin{equation}
\tan 2\theta = \frac{n_1n_2n_3(n^2_2-n^2_1)}{n^2_1n^2_2(1+n^2_3)-n^2_3(1-n^2_3)^2},
\label{equ316}
\end{equation}
where, if $\theta$ is a solution of \eq{equ316}, so is $\theta + \pi/2$.

\begin{figure}[htbp]
\centering
\vspace*{1mm} \rotatebox{0}{\resizebox{!}{7.5cm}{%
\includegraphics[scale=1]{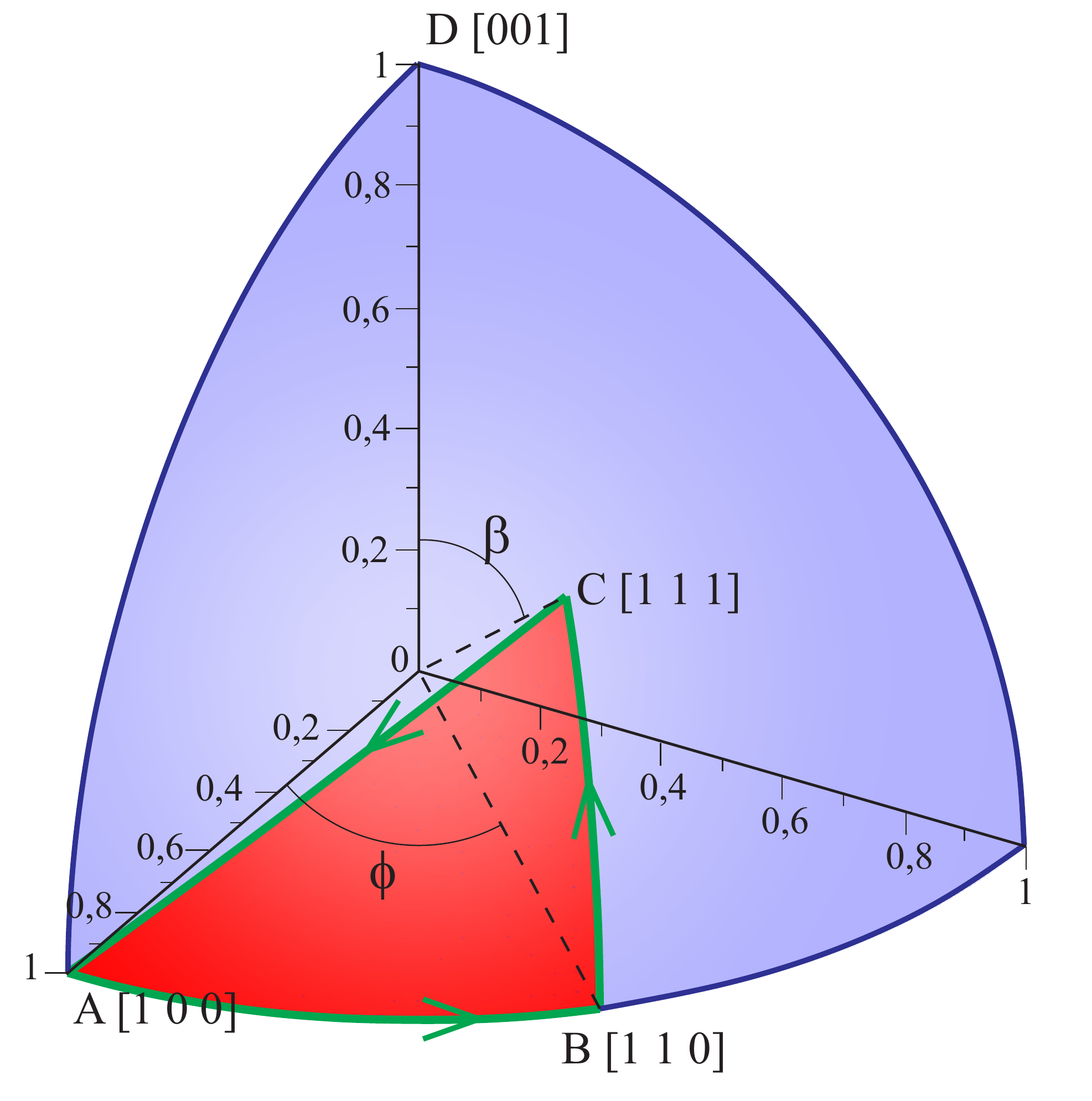}}}
 \caption{The irreducible spherical sector defined by the vertices corresponding to, $\textbf{n}=100,110$ and $111$. Note that the Poisson's ratios on the edges $\overline{C\!-\!D}$ and $\overline{C\!-\!A}$ are equal.}
\label{fig2_1_11}
\end{figure}

Following symmetry considerations the analysis can be reduced to the irreducible spherical sector delimited by directions $\textbf{n}=100,\,110,\,111$, which corresponds to $1/48th$ of the unit sphere surface as shown in Fig. \ref{fig2_1_11}, where A, B and C correspond to $\textbf{n}=100,110$ and $111$, respectively.
In \cite{Norris2006} it is shown that there are no stationary values of $\nu^*(\textbf{n},\textbf{m})$ inside the spherical sector of Fig. \ref{fig2_1_11}, but only on its boundary. Hence, we follow the local minimum and maximum Poisson's ratio when $\textbf{n}$ moves along the path $A\!-\!B\!-\!C\!-\!A$ visualized in green in Fig. \ref{fig2_1_11}, while the critical directions of $\textbf{m}$ making $\nu^*$ stationary are obtained from Eq. (\ref{equ316}).

\begin{figure}[htbp]
\centering
\vspace*{0mm} \rotatebox{0}{\resizebox{!}{9.5cm}{%
\includegraphics[scale=1]{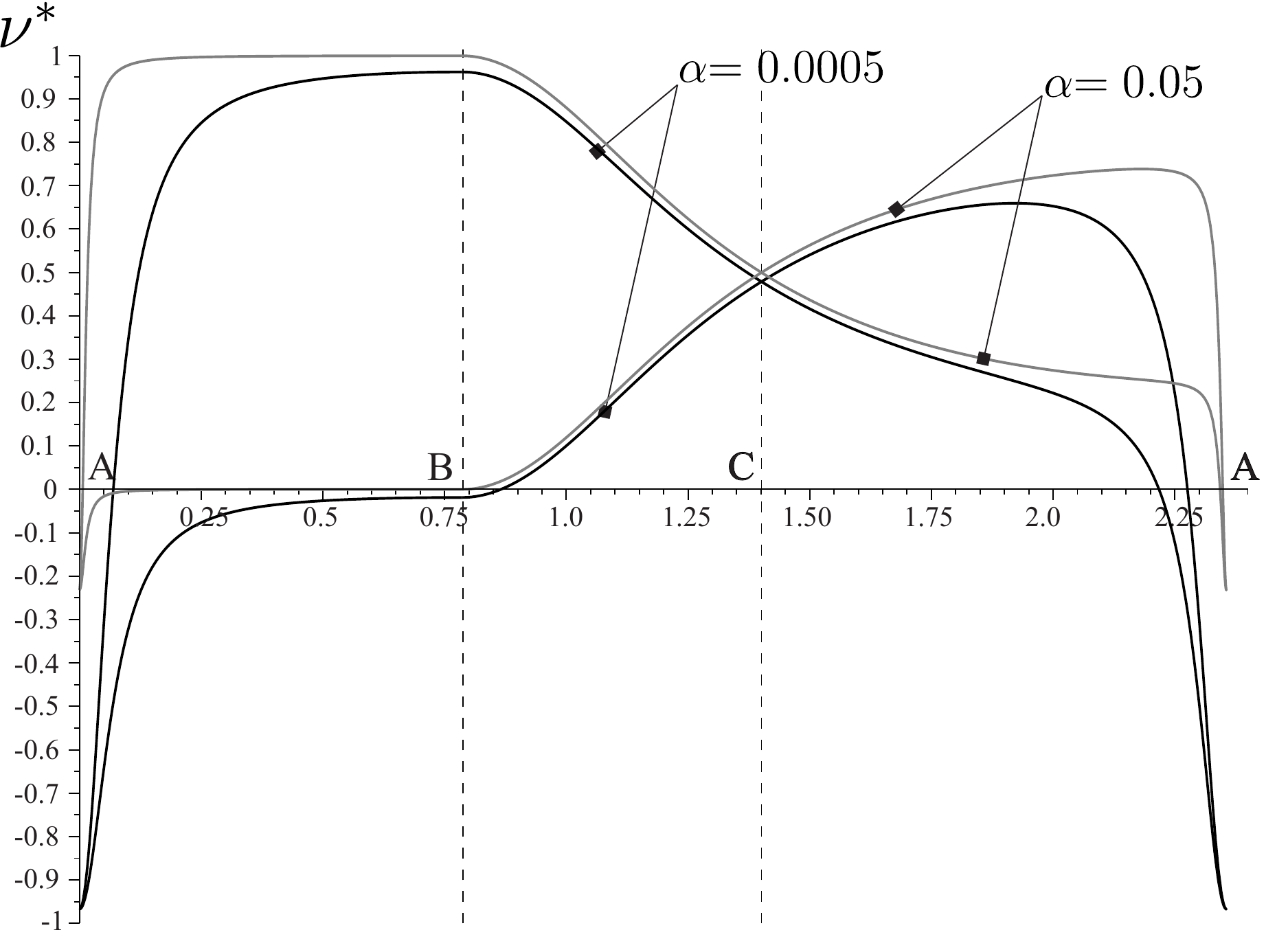}}}
 \caption{Maximum and minimum Poisson's ratio $\nu^*$ for different position of \textbf{n} along the path $A-B-C-A$ shown in Fig. \ref{fig2_1_11}. Results are given for $\alpha=0.0005$ (black lines) and $\alpha=0.05$ (grey lines). For $\alpha=0.0005$ the extreme values are $\nu^*_{min}=-0.967$ and $\nu^*_{max}=0.925$. For $\alpha=0.05$ the extreme values are $\nu^*_{min}=-0.232$ and $\nu^*_{max}=0.998$.  }
\label{fig2_1_12}
\end{figure}

We show in Fig. \ref{fig2_1_12} the local minima and maxima along the path $A\!-\!B\!-\!C\!-\!A$ among which it is possible to find the global minimum and maximum. Results are given for $\alpha=0.0005$ and $0.05$.
In both cases the global minimum is at the point A, where $\textbf{n}=  \,100$, and it is independent on the orthogonal direction $\textbf{m}=  \,0ab$, as can be easily checked from Eq. (\ref{equ312}). The global maximum is at the point B, with  $\textbf{n}=  \,110$ and $\textbf{m}=  \,1\bar{1} 0$, corresponding to $\theta=0$ in Eq. (\ref{equ316}).

In Fig. \ref{fig2_1_12} we also note the high degree of anisotropy of the lattice, the Poisson's ratio increases quite rapidly moving from the direction $\textbf{n}=100$. In the direction $\textbf{n}=1p0$, with $0\leq p \leq 1$, along $A\!-\!B$, the behaviour is always auxetic along specific transversal directions, while in the proximity of $\textbf{n}=111$ (point C) the behaviour is not auxetic.

For a generic stiffness ratio $\alpha$, the five possible candidates for the global minimum and maximum of $\nu^*(\textbf{n},\textbf{m})$ are the following \cite{Norris2006}:
\begin{eqnarray}
\nonumber
\nu^*_{001}\!=\!\nu^*(110,001)=-\frac{2S_{1122}}{S_{1111}+S_{1122}+2S_{1212}}, \\
\nonumber
\nu^*_{1\bar{1}0}\!=\!\nu^*(110,1\bar{1}0)=-\frac{S_{1111}+S_{1122}-2S_{1212}}{S_{1111}+S_{1122}+2S_{1212}},\\
\nonumber
\nu^*_0=\nu^*(100,-)=-\frac{S_{1122}}{S_{1111}}, \\
\nonumber
\nu^*_1=
\nu^*(11p_1,1\bar{1}0)=
-\frac{1}{4}\left(\frac{2S_{1111}+4S_{1122}-4S_{1212}}{S_{1111}+2S_{1122}+4S_{1212}}+
\frac{\sqrt{C_1}}{|S_{1111}+2S_{1122}+4S_{1212}|}\right),\\
\nu^*_2=\nu^*(11p_2,p_2p_2\bar{2})=
\frac{3S_{1212}}{S_{1111}+2S_{1122}+4S_{1212}}+
\frac{\sqrt{C_2}}{|S_{1111}+2S_{1122}+4S_{1212}|},
\label{equ317}
\end{eqnarray}
where
\begin{eqnarray}
\nonumber
p_1=\sqrt{\frac{\nu^*_1+1/2}{\nu^*_1-1/4}},\quad
p_2=\sqrt{\frac{\nu^*_2-3/2}{\nu^*_2-3/4}},\\
\nonumber
C_1=6[S_{1111}^2+3S_{1111}S_{1122}+2S_{1122}^2-2(S_{1111}+4S_{1122})S_{1212}],\\
C_2=\frac{3}{8}[3S_{1111}^2+11S_{1111}S_{1122}+10S_{1122}^2+2(3S_{1111}+4S_{1122})S_{1212}].
\label{equ318}
\end{eqnarray}
The entirety of possible cubic materials with positive definite strain energy are limited by the constraints
\begin{equation}
-1< \nu^*_{1\bar{1}0}< 1,\quad \mbox{and}\quad -\frac{1}{2}(1-\nu^*_{1\bar{1}0})< \nu^*_{001}<1-\nu^*_{1\bar{1}0}.
\label{equ319}
\end{equation}
These two relations define the interior of a triangle in the $(\nu^*_{001}, \nu^*_{1\bar{1}0})$ plane. The interior of the triangle represents the entirety of possible cubic materials with positive definite strain energy.

\begin {table}[!hcbt]
\begin{center}
\renewcommand\arraystretch{1.1}
\begin{tabular}{|c|c|c|c|c|}
\hline
\textbf{$\nu^*_{min}$} & \textbf{n}& \textbf{m}& \textbf{condition 1}& \textbf{condition 2}\\ \hline
$0<\nu^*_{001}$ & $110$ & $001$ & $\nu^*_{001}>0$ & $\nu^*_{1\bar{1}0}>\nu^*_{001}$\\[1 mm] \hline
$-1/2<\nu^*_{1\bar{1}0}$ & $110$ & $1\bar{1}0$ & $\nu^*_{1\bar{1}0}>-1/2$ & $\nu^*_{1\bar{1}0}<\nu^*_{001}$\\[1 mm] \hline
$-1<\nu^*_{0}$& $100$ & arbitrary & $\nu^*_{001}<0$ & $\nu^*_{1\bar{1}0}>\nu^*_{001}$\\[1 mm] \hline
$-\infty<\nu^*_{1}$& $11p_1$ & $1\bar{1}0$ & $\nu^*_{1\bar{1}0}<-1/2$ & $\nu^*_{1\bar{1}0}<\nu^*_{001}$\\[1 mm] \hline
\end{tabular}
\caption[Global minimum of Poisson's ratio for cubic material]{Global minimum $\nu^*_{min}$ of Poisson's ratio for cubic material based on the values $\nu^*_{001}$ and $\nu^*_{1\bar{1}0}$}
\label{Table2_1}
\end{center}
\end {table}

\begin {table}[!hcbt]
\begin{center}
\renewcommand\arraystretch{1.1}
\begin{tabular}{|c|c|c|c|c|}
\hline
\textbf{$\nu^*_{max}$} & \textbf{n}& \textbf{m}& \textbf{condition 1}& \textbf{condition 2}\\ \hline
$\nu^*_{1}<-1/2$ & $11p_1$ & $1\bar{1}0$ & $\nu^*_{1\bar{1}0}<-1/2$ & $\nu^*_{1\bar{1}0}>\nu^*_{001}$\\[1 mm] \hline
$\nu^*_{0}<0$ & $100$ & $arbitrary$ & $\nu^*_{001}<0$ & $\nu^*_{1\bar{1}0}<\nu^*_{001}$\\[1 mm] \hline
$\nu^*_{1\bar{1}0}<1$& $110$ & $1\bar{1}0$ & $\nu^*_{1\bar{1}0}>-1/2$ & $\nu^*_{1\bar{1}0}>\nu^*_{001}$\\[1 mm] \hline
$\nu^*_{001}<3/2$& $110$ & $001$ & $0<\nu^*_{001}<3/2$ & $\nu^*_{1\bar{1}0}<\nu^*_{001}$\\[1 mm] \hline
$\nu^*_{2}<\infty$& $11p_2$ & $p_2p_2\bar{2}$ & $\nu^*_{001}>3/2$ & $$\\[1 mm] \hline
\end{tabular}
\caption[Global maximum of Poisson's ratio for cubic material]{Global maximum $\nu^*_{max}$ of Poisson's ratio for cubic material}
\label{Table2_2}
\end{center}
\end {table}

\begin{figure}[!htcb]
\centerline{
	\begin{tabular}{c}
         	 	\begin{tabular}{c@{\hspace{0.5pc}}c@{\hspace{0.5pc}}c}
             		\includegraphics[width=0.5\columnwidth]{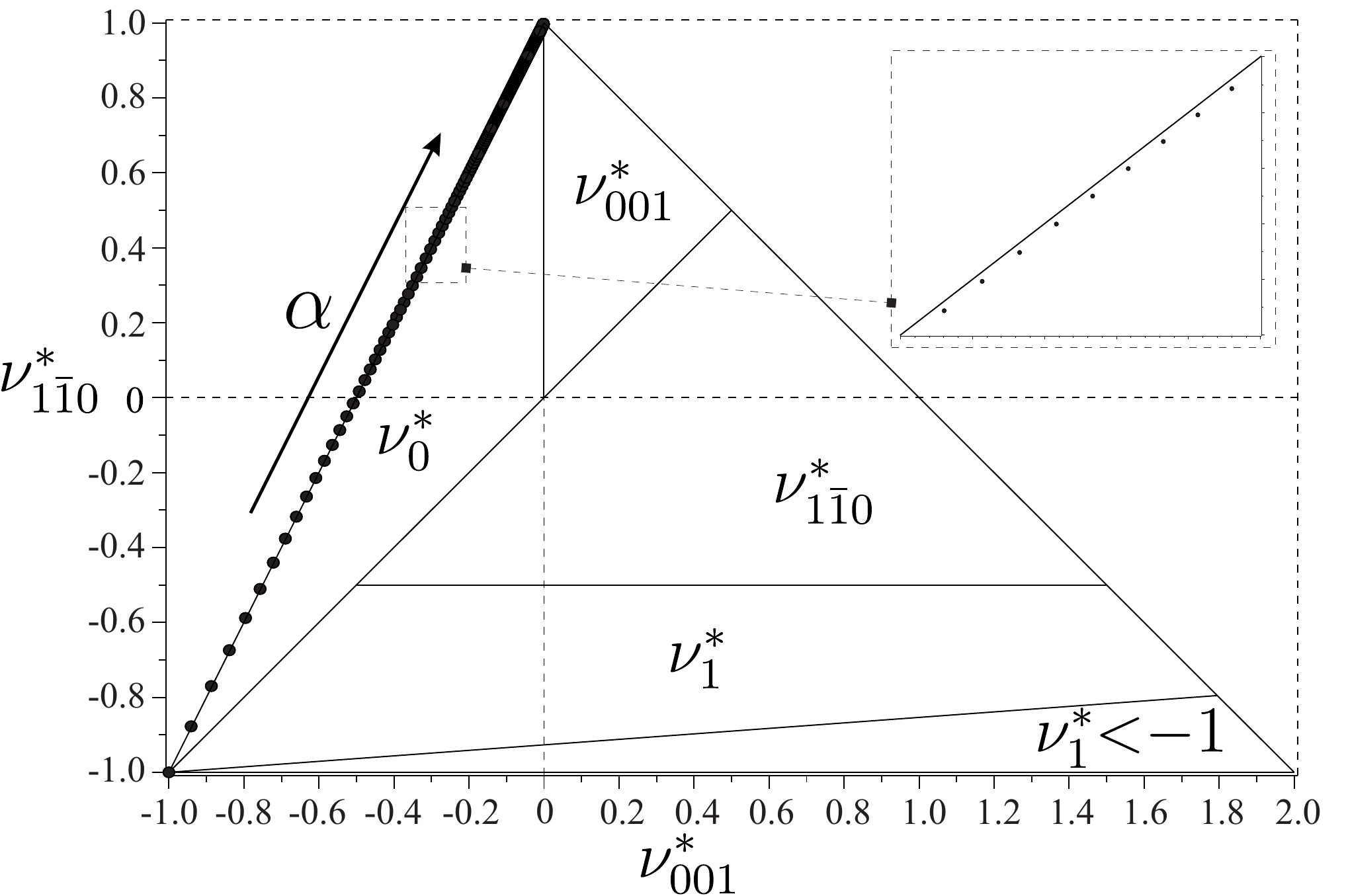} &
             		\includegraphics[width=0.5\columnwidth]{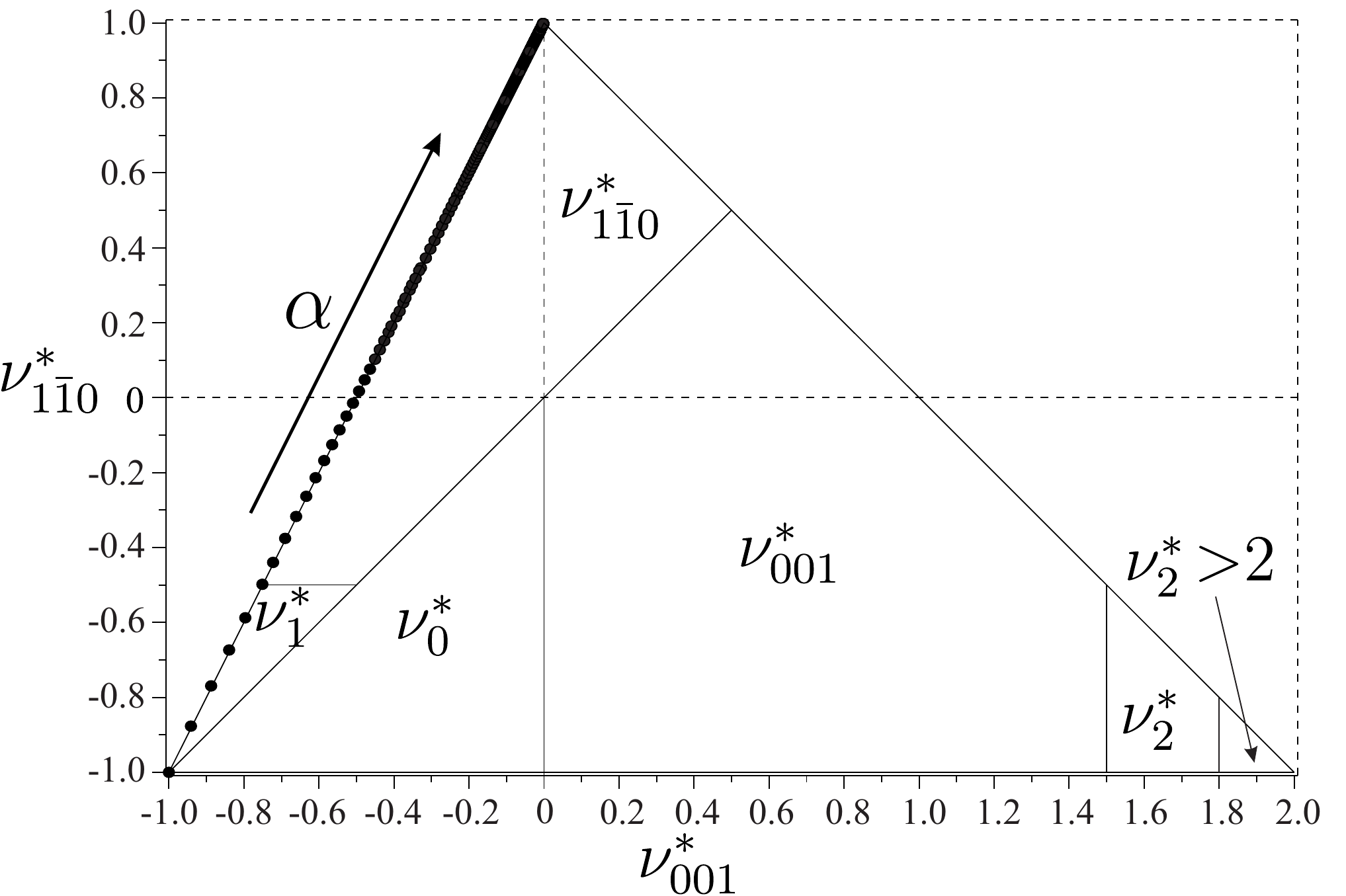}\\
         			(a) & (b)
         		\end{tabular} \\
                  \end{tabular}
       }
\caption{Global extrema of the Poisson's ratio $\nu^*(\textbf{n},\textbf{m})$ for a cubic material in the $(\nu^*_{001},\nu^*_{1\bar{1}0})$ plane. The dots represent the values of the cubic lattice for different values of the stiffness ratio $\alpha$. (a) Global minimum $\nu^*_{min}$. (b) Global maximum $\nu^*_{max}$.}
\label{fig2_1_13}
\end{figure}

Tables \ref{Table2_1} and \ref{Table2_2} list the values of the global minimum $\nu^*_{min}$ and the global maximum $\nu^*_{max}$, respectively, for all possible combinations of elastic parameters.
Each row of the Tables \ref{Table2_1} and \ref{Table2_2} identifies a sector in the triangles shown in Fig. \ref{fig2_1_13} and defined by Eq. (\ref{equ319}). These sectors define the global extrema for every point within the triangles, so to identify them we need to know the two quantities $\nu^*_{001}$ and $\nu^*_{1\bar{1}0}$, from which it is possible to go back to the values sought. In Fig. \ref{fig2_1_13} we show the values of $\nu^*_{001}$ and $\nu^*_{1\bar{1}0}$ for our lattice, each dot representing a different value of $\alpha$. For $\alpha\to 0$, $(\nu^*_{001}, \nu^*_{1\bar{1}0})\to(-1,-1)$, where $K^*/\mu^*\to0$, while for $\alpha\to\infty$, $(\nu^*_{001}, \nu^*_{1\bar{1}0})\to(0,1)$, where $K^*/\mu^*\to \infty$.
For finite values of the stiffness ratio $\alpha$ the effective behavior is such that the corresponding points in the $(\nu^*_{001}, \nu^*_{1\bar{1}0})$ plane are close to the constitutive stability limit $\nu^*_{001}+\frac{1}{2}(1-\nu^*_{1\bar{1}0})=0$, where ${\hat{\mu}}^*\to\infty$, but inside the triangle, as detailed in the inset of Fig. \ref{fig2_1_13}a.
The minimum Poisson's ratio for different $\alpha$ is always given by $\nu^*_{0}$, while the maximum is $\nu^*_1$ for $\alpha<0.326\cdot 10^{-5}$ and $\nu^*_{1\bar{1}0}$ for $\alpha>0.326\cdot 10^{-5}$.

\begin{figure}[!htcb]
\centerline{
	\begin{tabular}{c}
         	 	\begin{tabular}{c@{\hspace{2.5pc}}c@{\hspace{0.5pc}}c}
             		\includegraphics[height=7cm]{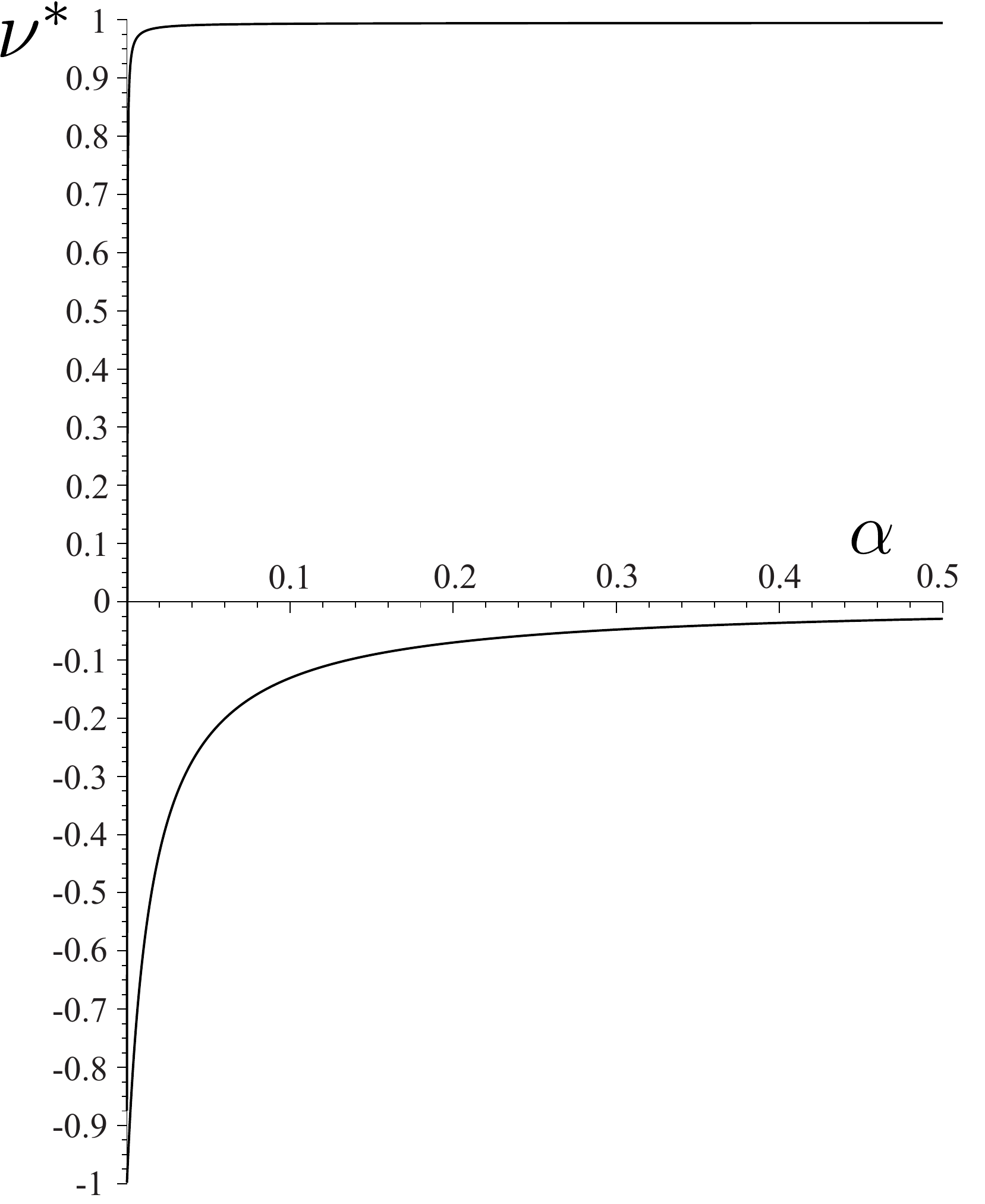} &
             		\includegraphics[height=7cm]{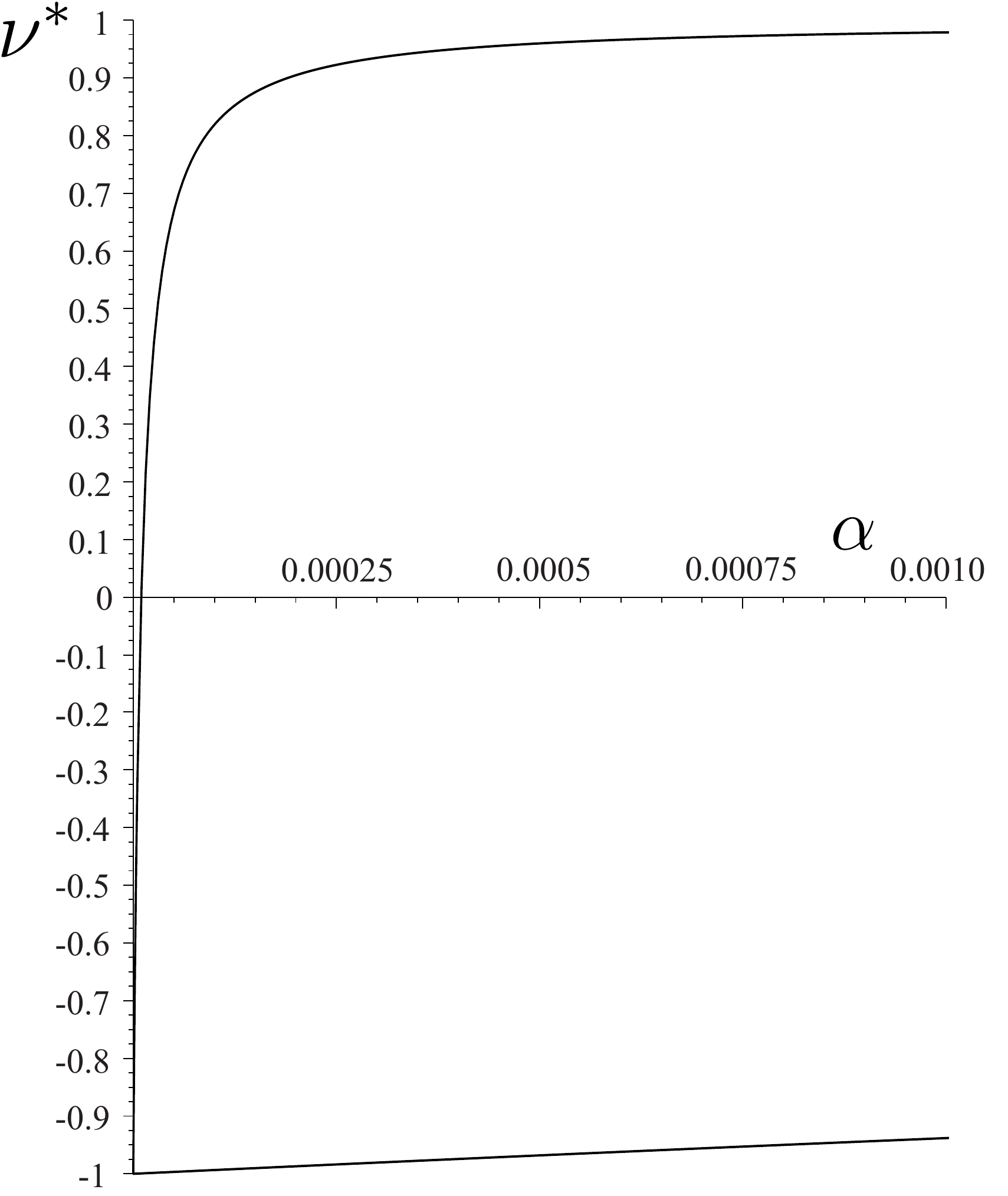}\\
         			(a) & (b)
         		\end{tabular} \\
                  \end{tabular}
       }
\caption{Global minimum and maximum Poisson's ratio $\nu^*$ as a function of the dimensionless stiffness ratio $\alpha$. (a) $\alpha\in(0,0.5)$. (b) $\alpha\in(0,0.01)$. }
\label{fig2_1_14}
\end{figure}

In Fig. \ref{fig2_1_14} we show the global minimum and maximum of the Poisson's ratio 
as a function of the dimensionless parameter $\alpha$. The behaviour of the lattice is always auxetic when loaded along one of the principal directions $\textbf{n}=(100),(010),(001)$ of the lattice, as shown by the minimum $\nu^*_0$ and also in Fig. \ref{fig2_1_9}. Non uniform omni-directional negative Poisson's ratio is achieved in the narrow interval $0\le\alpha< 10^{-5}$, indeed decreasing the stiffness ratio $\alpha$ to $0$ the global minimum and maximum of $\nu^*$ tend to $-1$, which is a stability limit.
Increasing $\alpha$ the global maximum tends rapidly to $1^-$ and the global minimum tends to $0^-$.

In conclusion, the cubic lattice presents omni-directional negative Poisson's ratio, which varies with \textbf{n} and \textbf{m}, only in a limited range of $\alpha$; in general, for every $\alpha$, the behavior is  auxetic when the unidirectional stress is applied along one of the principal direction of the lattice. The stability limit $\nu^*\to -1$ can be achieved when $\alpha\to 0$, where, in turn, the structure becomes isotropic.

\section{Isotropic auxetic lattice}
\label{Sect3}

\begin{figure}[!htcb]
\centerline{
	\begin{tabular}{c}
         	 	\begin{tabular}{c@{\hspace{-1.0pc}}c@{\hspace{-1.0pc}}c}
             		\includegraphics[height=4.5cm]{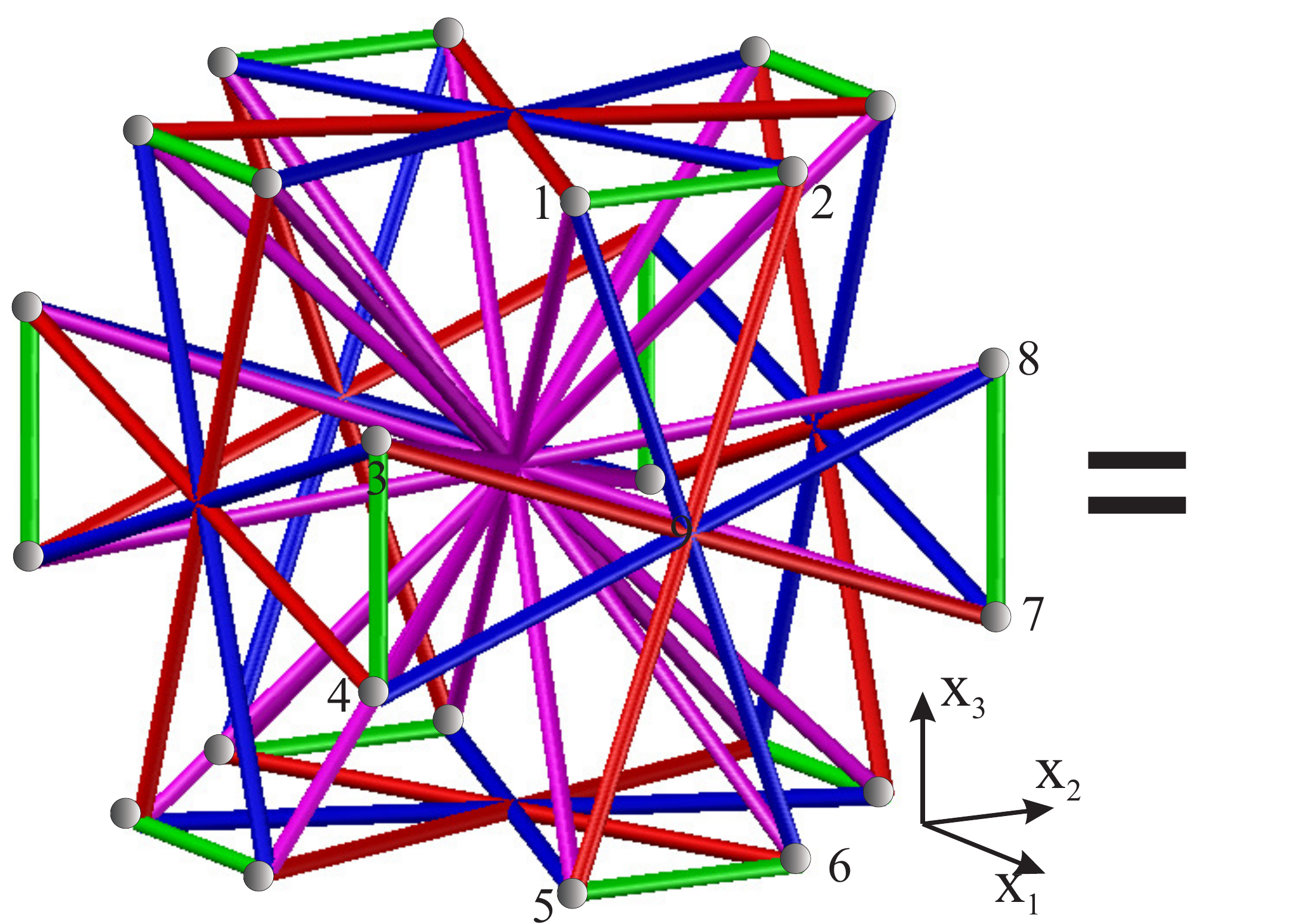} &
                    \includegraphics[height=4.5cm]{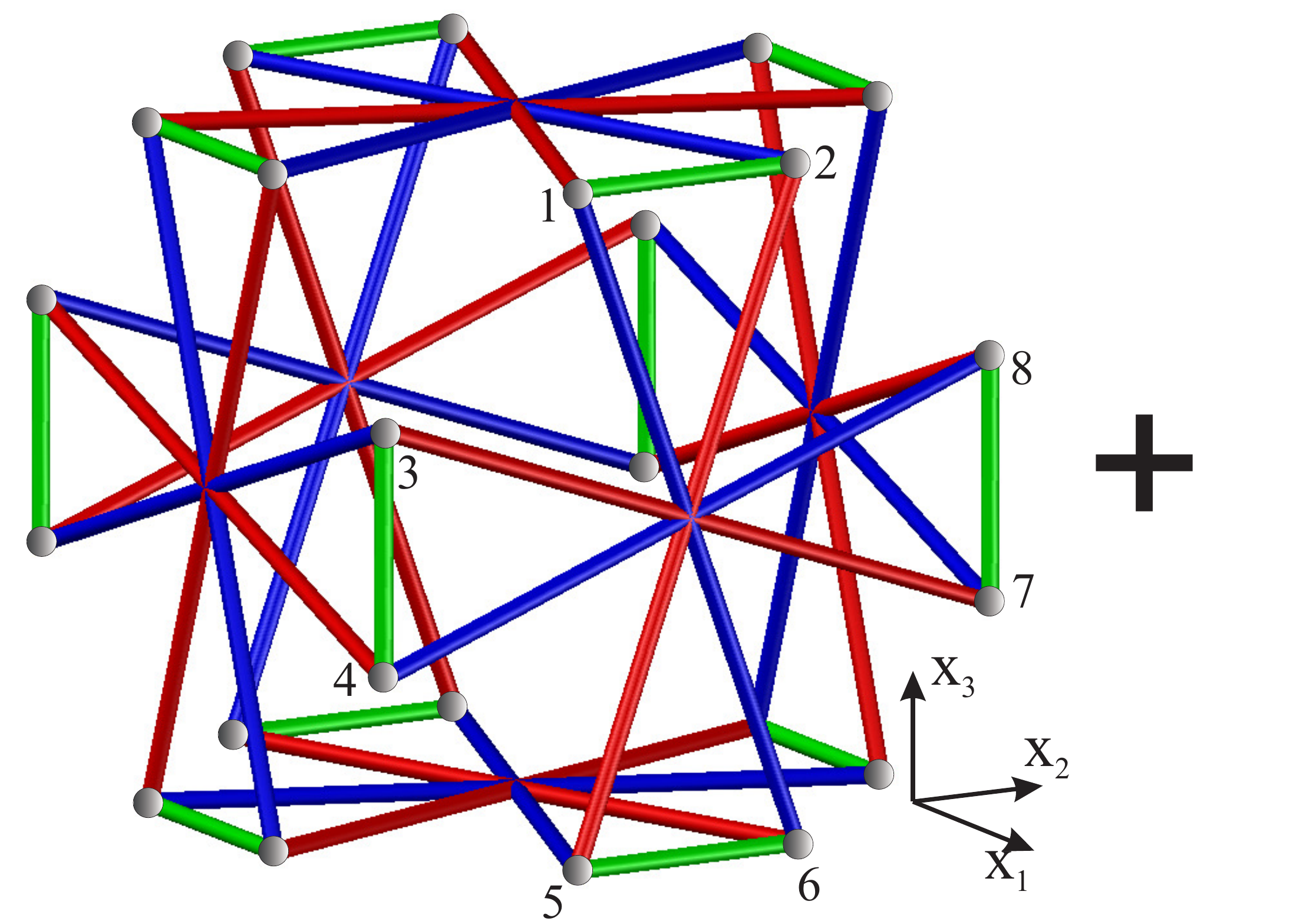} &
             		\includegraphics[height=4.5cm]{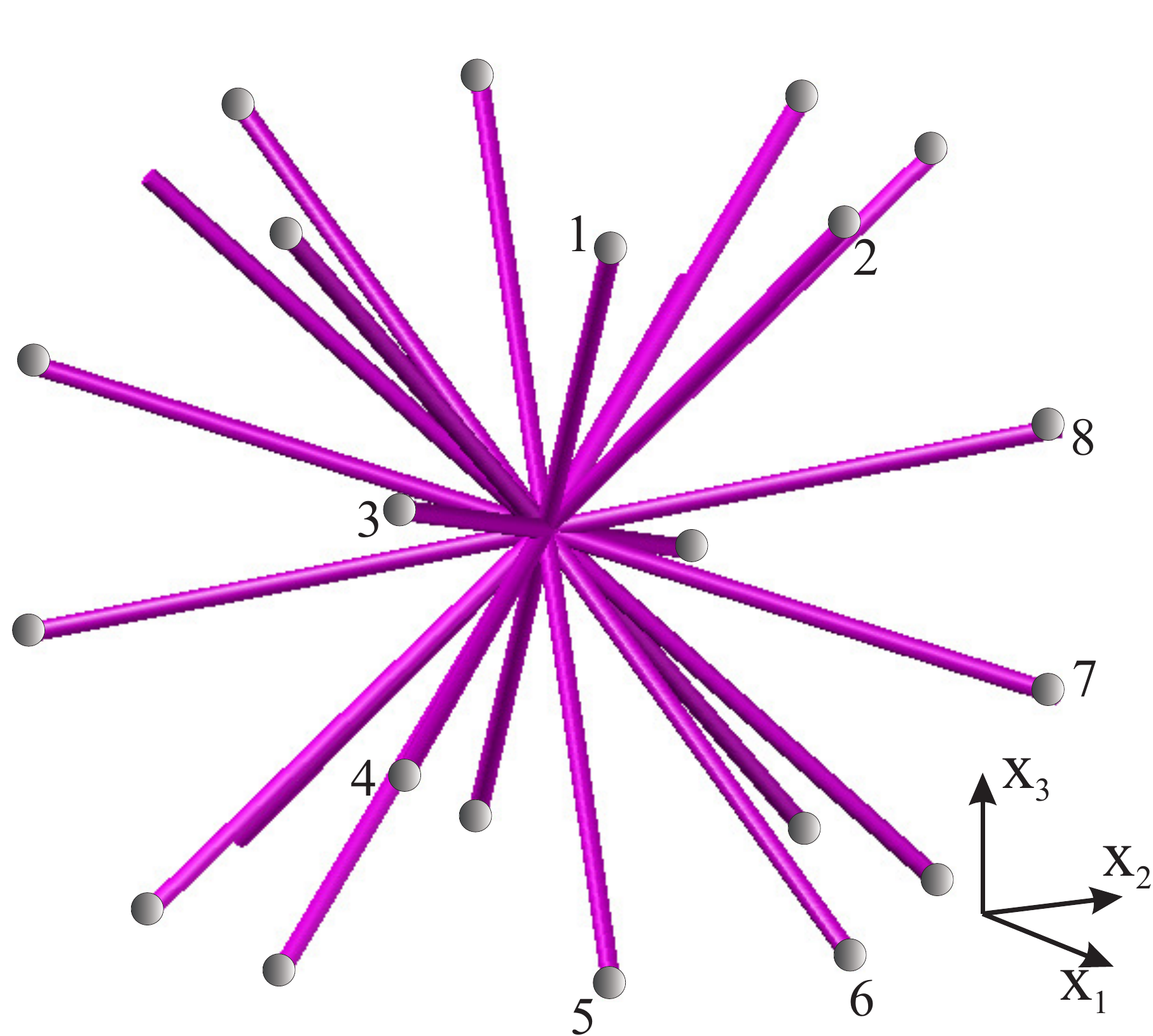}\\
         			(a) & (b)& (c)
         		\end{tabular} \\
                  \end{tabular}
       }
\caption{(a) Three-dimensional unit cell of the isotropic lattice. (b) Previously introduced cubic lattice (depicted in blue and red for the cross-shaped elements and in green for the truss elements). (c) System of additional diagonal elements (depicted in magenta). Numbers identify lattice points.}
\label{fig2_2_1}
\end{figure}

In the previous section we presented a cubic lattice, where the Poisson's ratio is negative only with respect to some directions $\textbf{n}$ and $\textbf{m}$, as shown in Fig. \ref{fig2_1_12}-\ref{fig2_1_14}. Here, we modify the previous micro-structure and we present a type of three-dimensional lattice that can exhibit isotropic negative Poisson's ratio.
In Fig. \ref{fig2_2_1} we show the modified lattice, which is obtained introducing diagonal beams depicted in magenta and adding a lattice point in the center of the unit cell to the eight corner points. Each diagonal beam is mutually constrained to have the same displacement at the central point where a hinge is introduced; in addition each beam is also constrained by internal hinges at its ends. Doing the same similarity with the crystal system used previously, such a lattice structure is a body-centered cubic system (cI). We indicate with $\eta$ the ratio between the cross-sectional areas of the just introduced diagonal beams $A_d$ and that of the arms of the cross-shaped elements $A_c$.

\begin{figure}[!htcb]
\centerline{
	\begin{tabular}{c}
         	 	\begin{tabular}{c@{\hspace{0.3pc}}c@{\hspace{0.3pc}}c@{\hspace{0.3pc}}c}
             		\includegraphics[width=0.31\columnwidth]{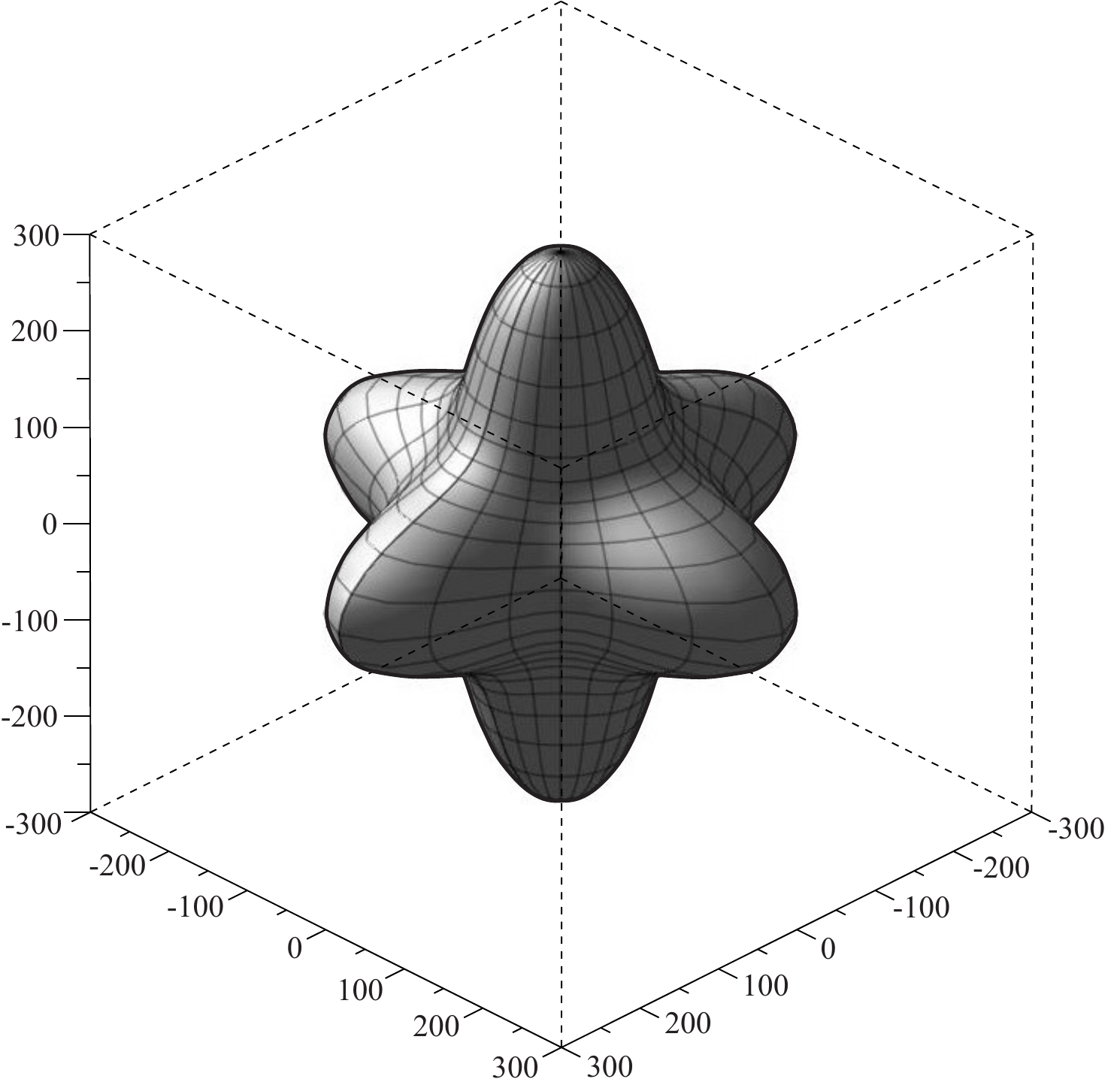} &
             		\includegraphics[width=0.31\columnwidth]{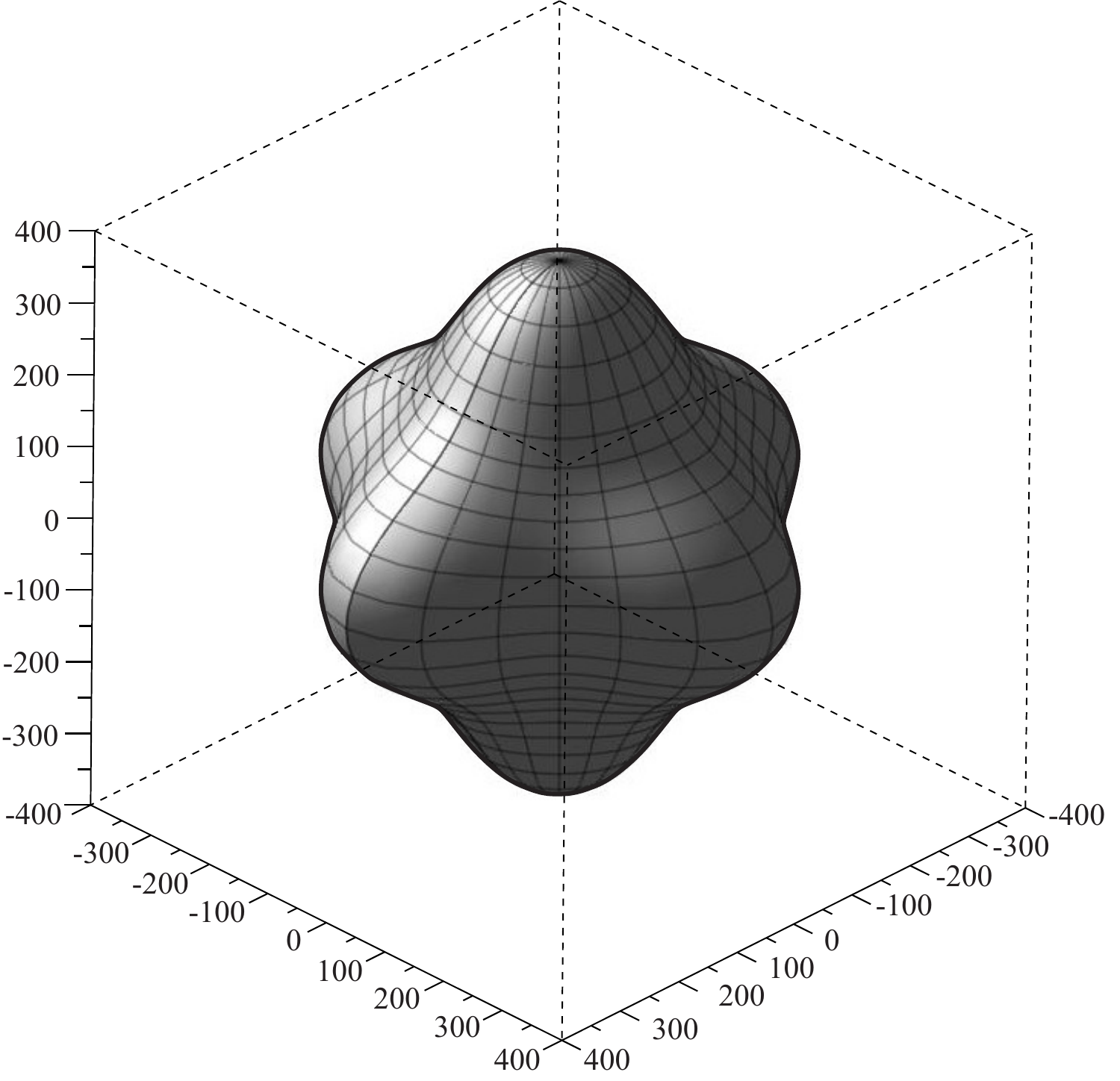} &
                    \includegraphics[width=0.31\columnwidth]{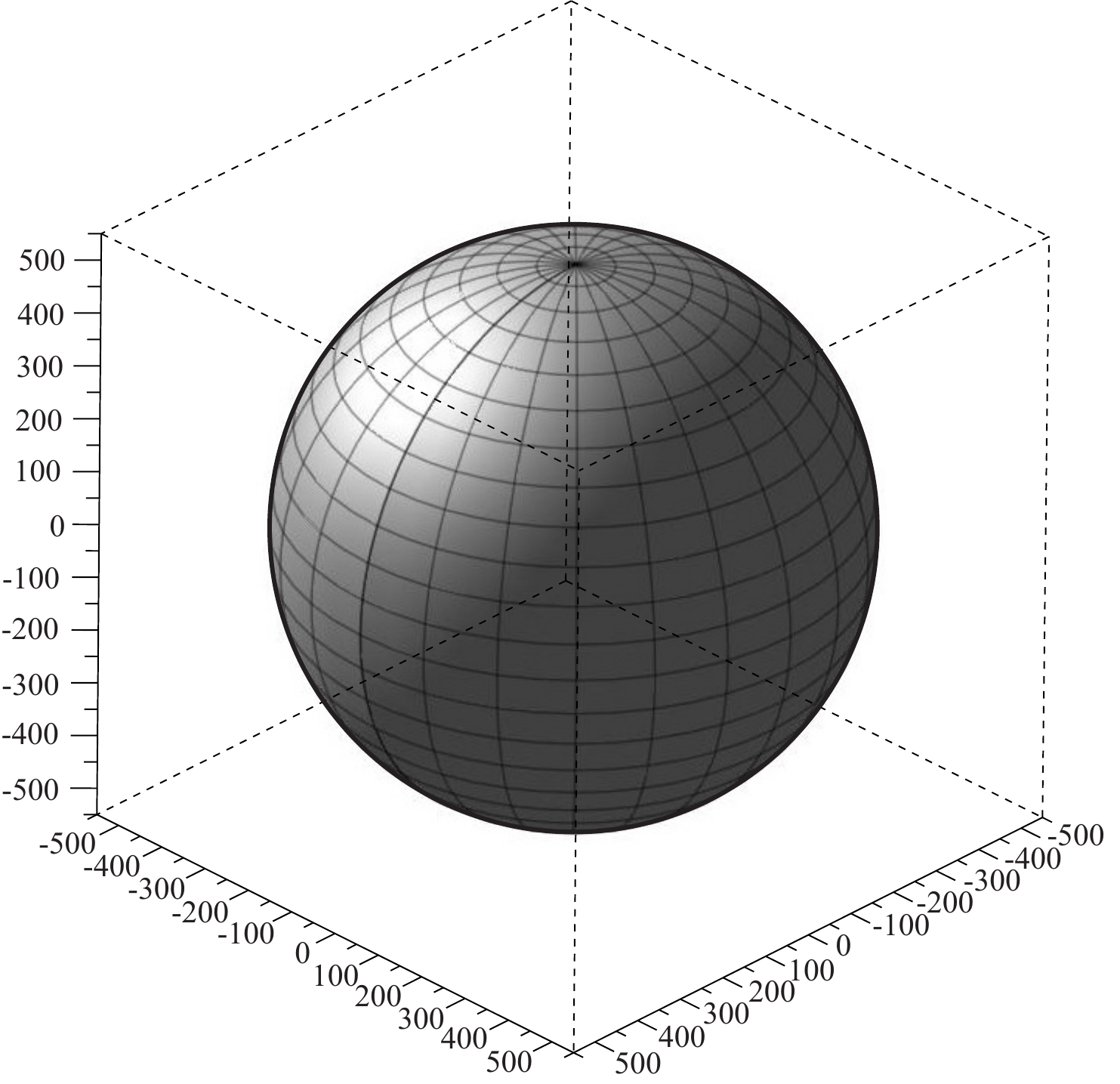} \\
                    (a) & (b) & (c)\\
                    \includegraphics[width=0.31\columnwidth]{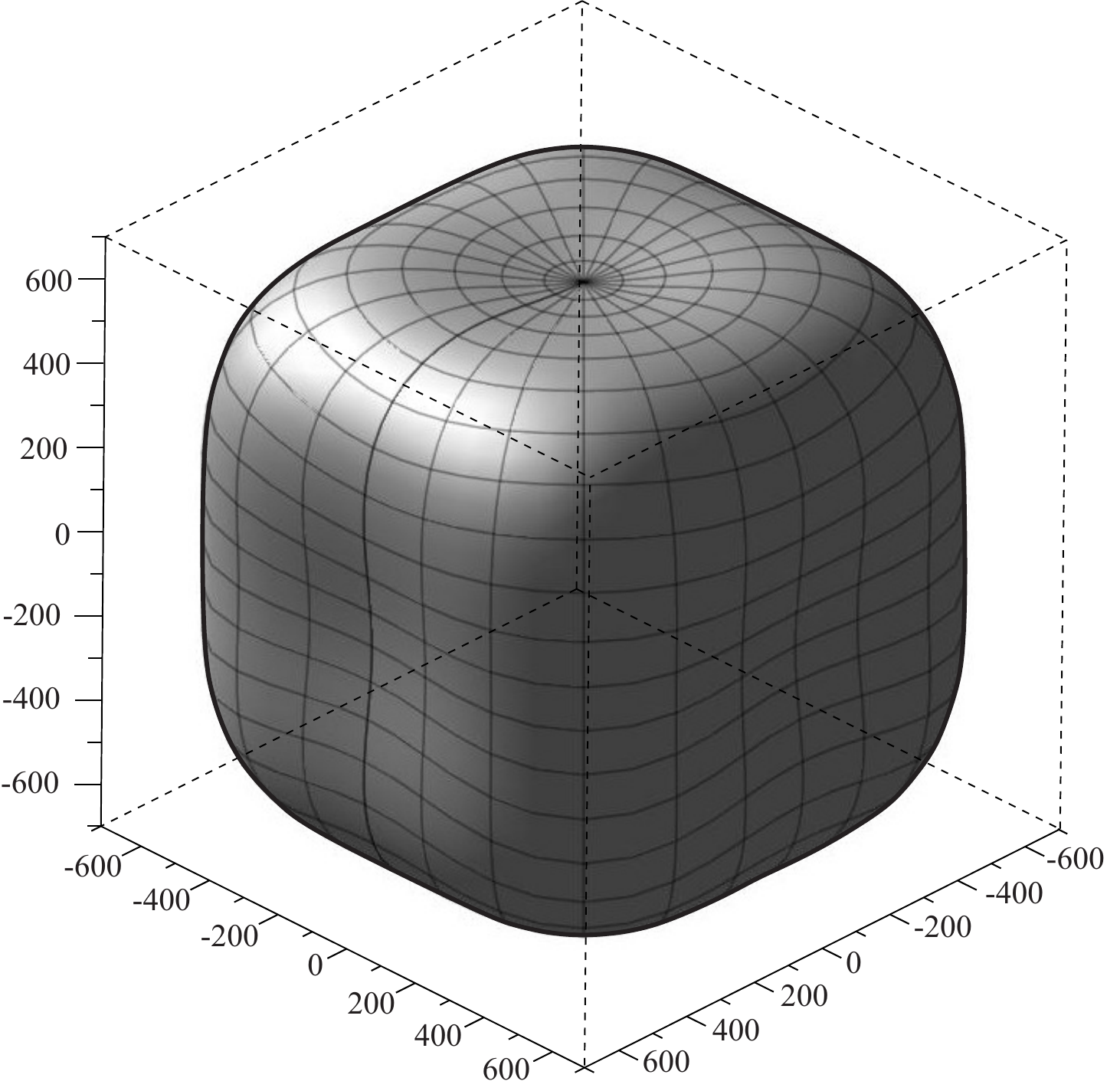} &
             		\includegraphics[width=0.31\columnwidth]{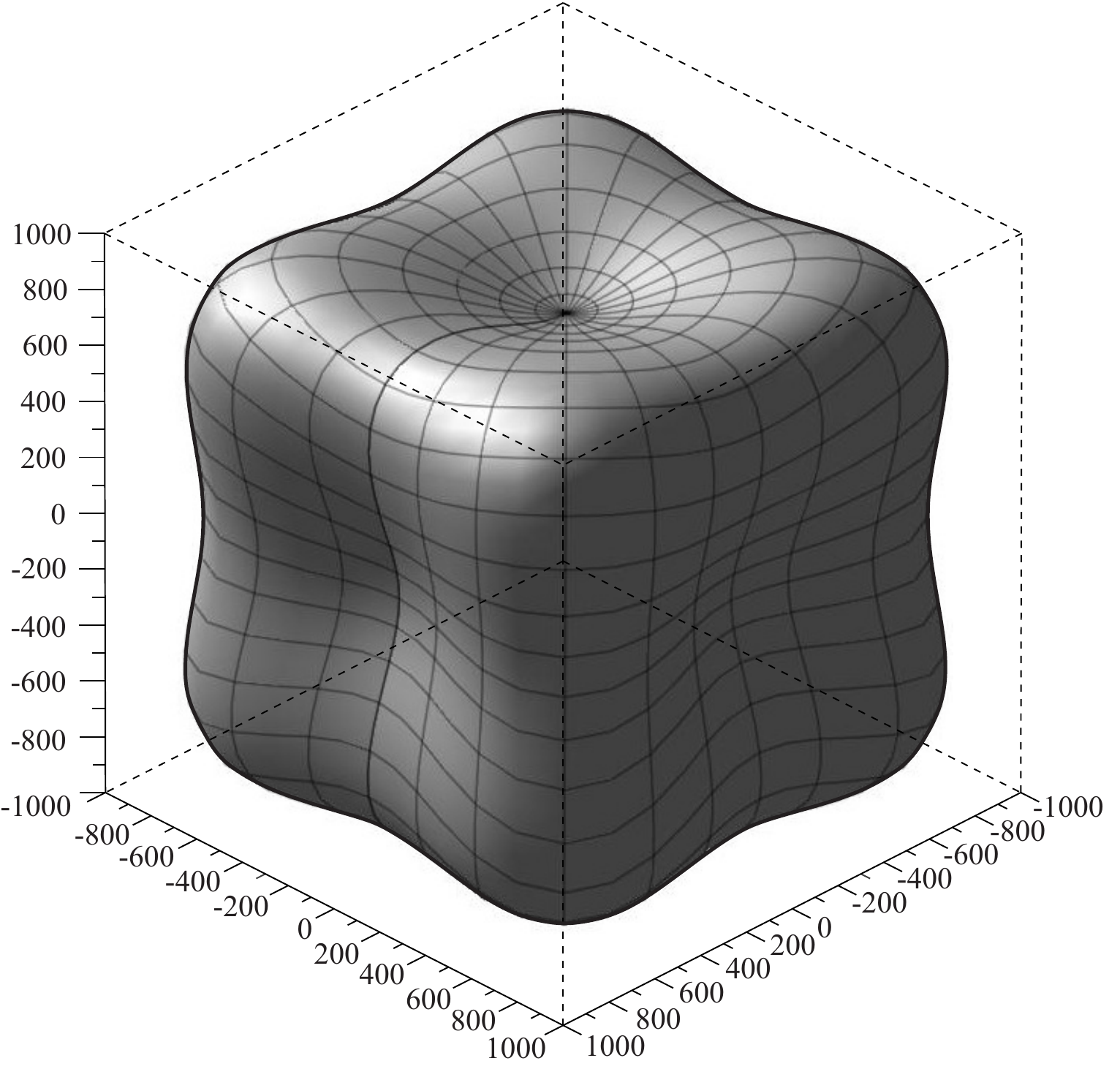} &
                    \includegraphics[width=0.31\columnwidth]{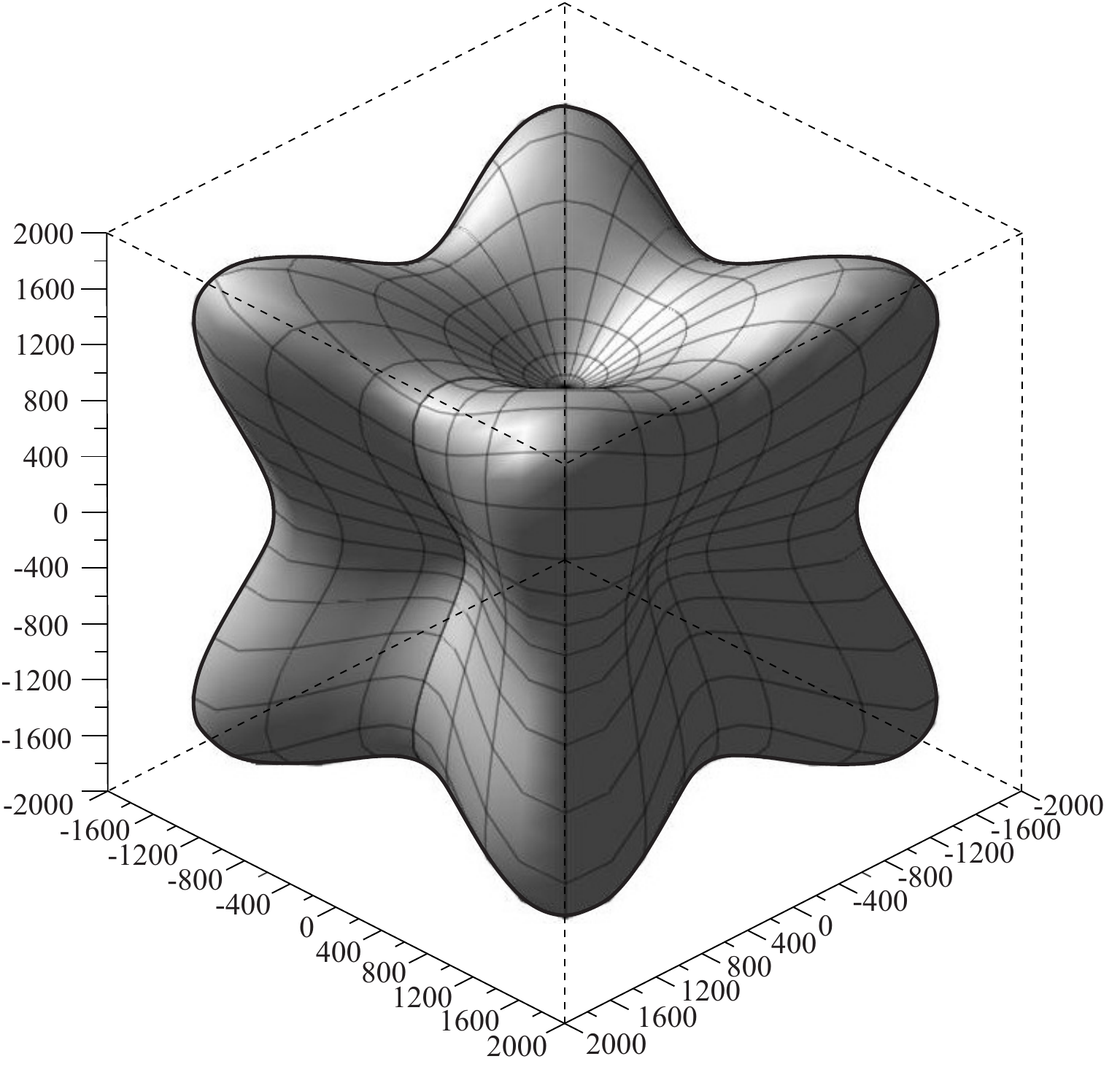} \\
         			(d) & (e) & (f)
         		\end{tabular} \\
                  \end{tabular}
       }
\caption{Directional dependence of the homogenized Young's modulus $E^*$ for six different cross-shaped ratio $\eta$ and for $\alpha=0.00515$. (a) $\eta=0.5$. (b) $\eta=1$. (c) $\eta=2.386$. (d) $\eta=5$. (e) $\eta=10$. (f) $\eta=100$.}
\label{fig2_2_4}
\end{figure}

The effective constitutive parameters are analysed as a function of the non-dimensional stiffness ratios $\alpha$ and $\eta$.
The directional dependance of the Young's modulus $E^*$ is shown in Fig. \ref{fig2_2_4}. Results are given for $\alpha=0.00515$ and each plot corresponds to increasing values of the ratio $\eta$. The comparative analysis indicates a transition from `weak shear resistant' structures, with Zener's anisotropy factor $\beta_{cub}<1$ to `dominating shear resistant' structures, with $\beta_{cub}>1$. At the transition point the Young's modulus is a sphere in the $(x_1x_2x_3)$ space, indicating isotropic effective behavior, corresponding also to $\beta_{cub}=1$.

\begin{figure}[!htbp]
\centerline{
         \begin{tabular}{c@{\hspace{2.5pc}}c@{\hspace{0.5pc}}c}
             		\includegraphics[height=7cm]{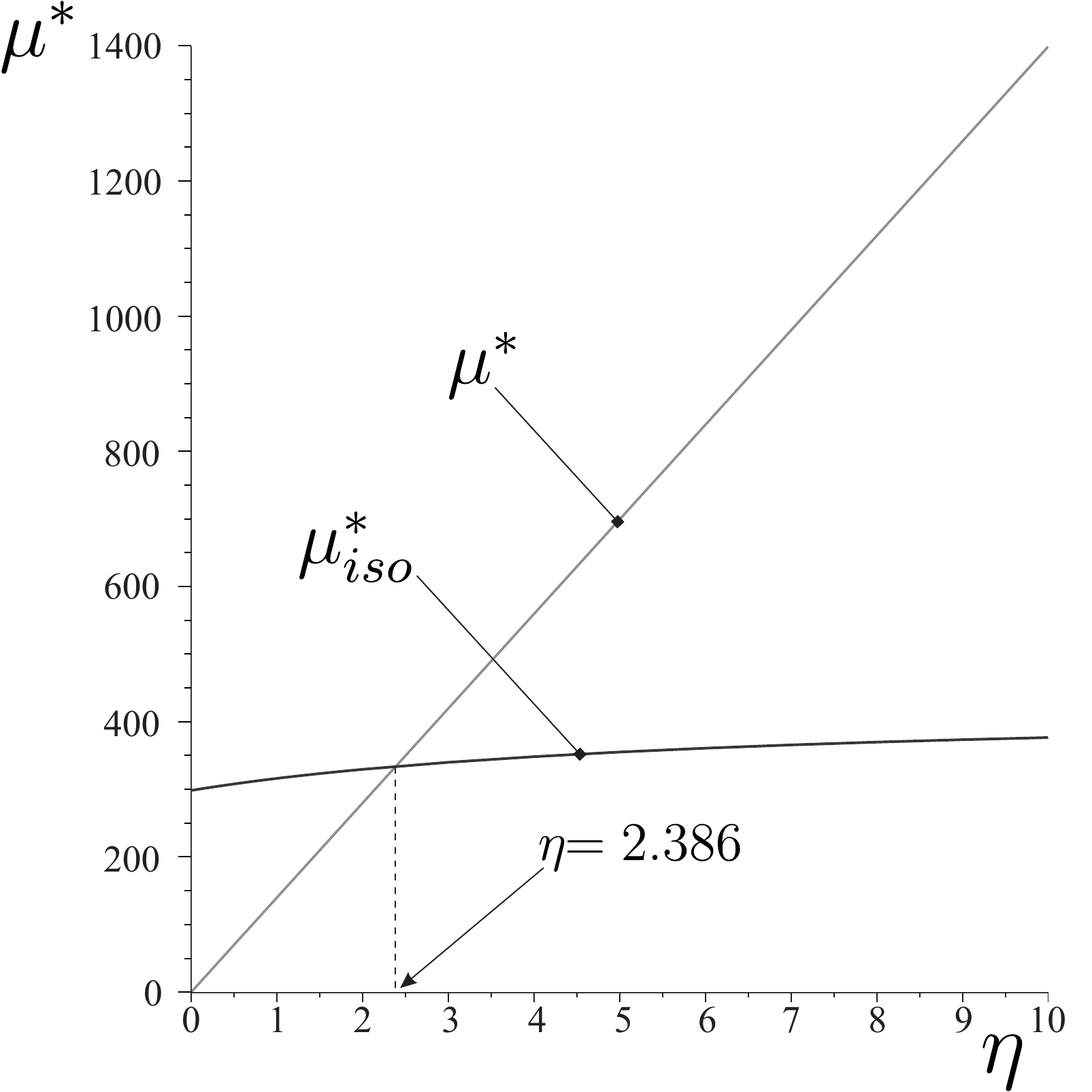} &
             		\includegraphics[height=7cm]{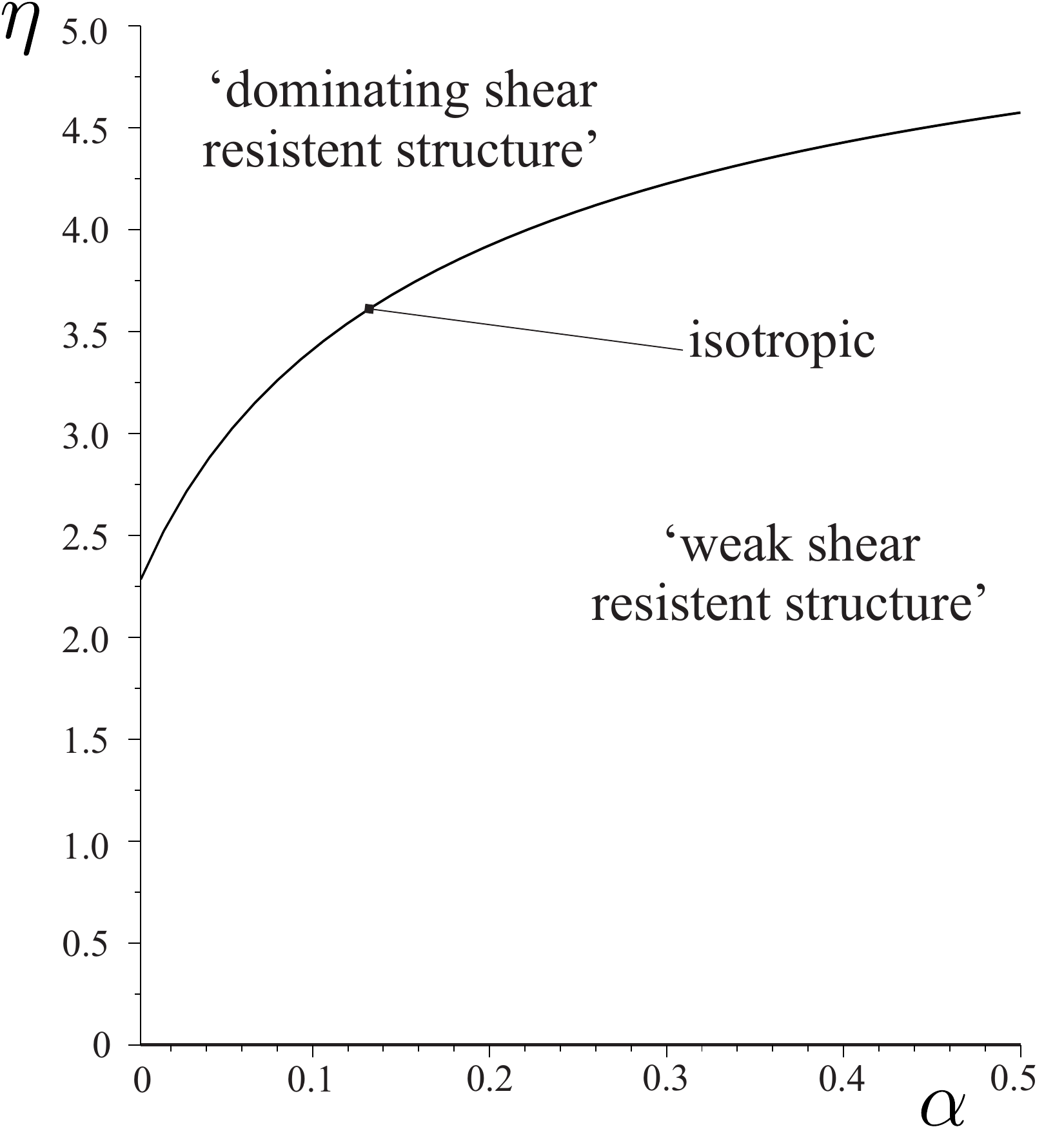}\\
         			(a) & (b)
         \end{tabular}
         }
\caption{(a) Effective shear moduli $\mu^*$ and $\mu^*_{iso}=E_{100}^*/(2(1+\nu^*_0))$ of the lattice as a function of the non-dimensional cross-section ratio $\eta$. Results correspond to $\alpha=0.00515$. (b) Non-dimensional cross-sectional ratio $\eta$ as a function of non-dimensional stiffness ratio $\alpha$ in order to have isotropic behaviour.}
\label{fig2_2_5}
\end{figure}

To identify the numerical value for which the ratio $\eta$ gives isotropic behavior at each $\alpha$, we compare the shear modulus $\mu^*_{100}=\mu^*$, see Eq. (\ref{eqn203}), with the shear modulus
$\mu^*_{iso}=E_{100}^*/(2(1+\nu^*_0))$ that the lattice would have in case of isotropic behavior. An example for $\alpha=0.00515$ is given in Fig. \ref{fig2_2_5}a; the intersection point at $\eta=2.386$ identifies the value for which the lattice has the desired isotropic behavior. Clearly, this corresponds to the value of $\eta$ for which $E^*_{100}=E^*_{110}=E^*_{111}$. The monotonic dependance of the cross-section ratio $\eta$ on the stiffness ratio $\alpha$ in order to obtain isotropic behaviour is reported in Fig. \ref{fig2_2_5}b. At these values the effective behavior is isotropic; therefore the Poisson's ratio does not depend on the directions \textbf{n} and \textbf{m} and, when plotted along the path $A-B-C-A$ as in Fig. \ref{fig2_1_12}, the local and global maxima and minima coincide and do not depend on the position within the path.

\begin{figure}[!htbp]
\centering
\vspace*{1mm} \rotatebox{0}{\resizebox{!}{8.0cm}{%
\includegraphics[scale=1]{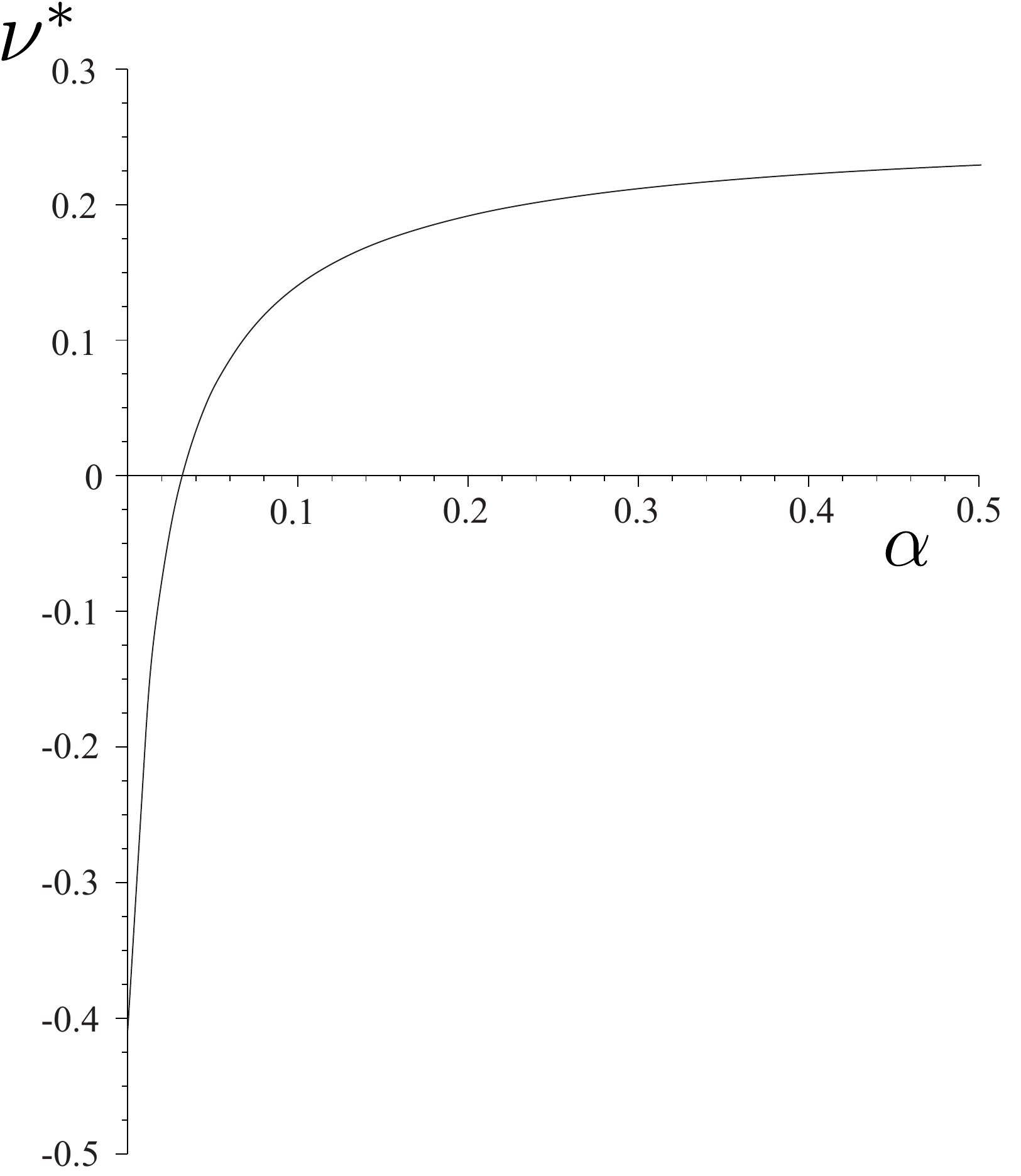}}}
 \caption{Poisson's ratio $\nu^*$ as a function of $\alpha$ in case of isotropic behaviour.}
\label{fig2_2_10}
\end{figure}

The isotropic effective Poisson's ratio as a function of the stiffness ratio $\alpha$ is shown in Fig. \ref{fig2_2_10}. In addition to the property of isotropy, we note that the interval in which the behavior is omni-directionally auxetic is strongly increased with respect to the cubic case in Section \ref{Cubic Section}, where $\eta=0$; more precisely the Poisson's ratio is negative for $\alpha<0.0325$. In contrast, it is not possible to reach the lower and upper limits $\nu^*=-1$ and $\nu^*=0.5$, respectively.

\begin{figure}[!htcb]
\centerline{
	\begin{tabular}{c}
         	 	\begin{tabular}{c@{\hspace{0.5pc}}c@{\hspace{0.5pc}}c}
             		\includegraphics[height=4.5cm]{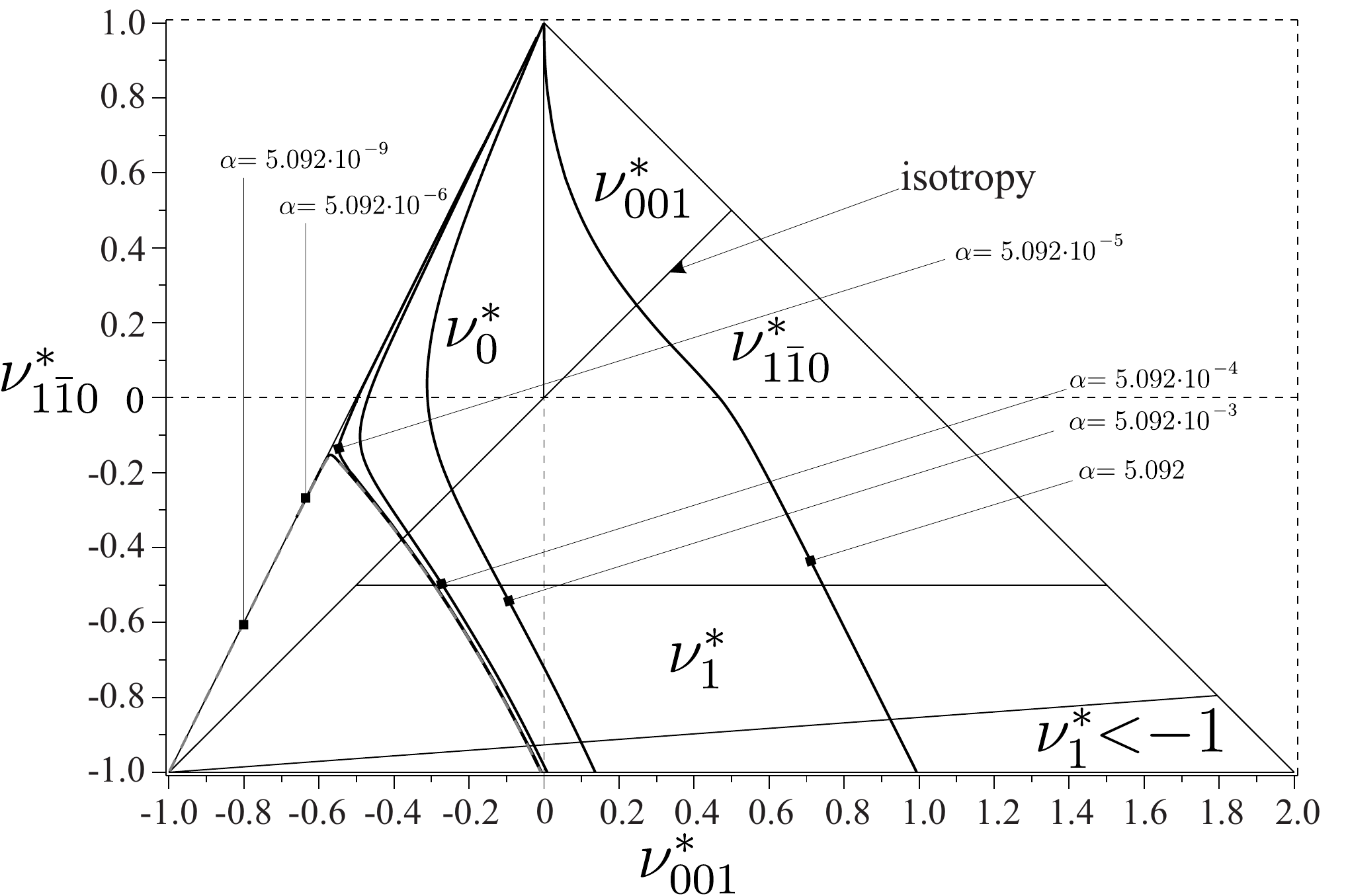}&
             		\includegraphics[height=4.5cm]{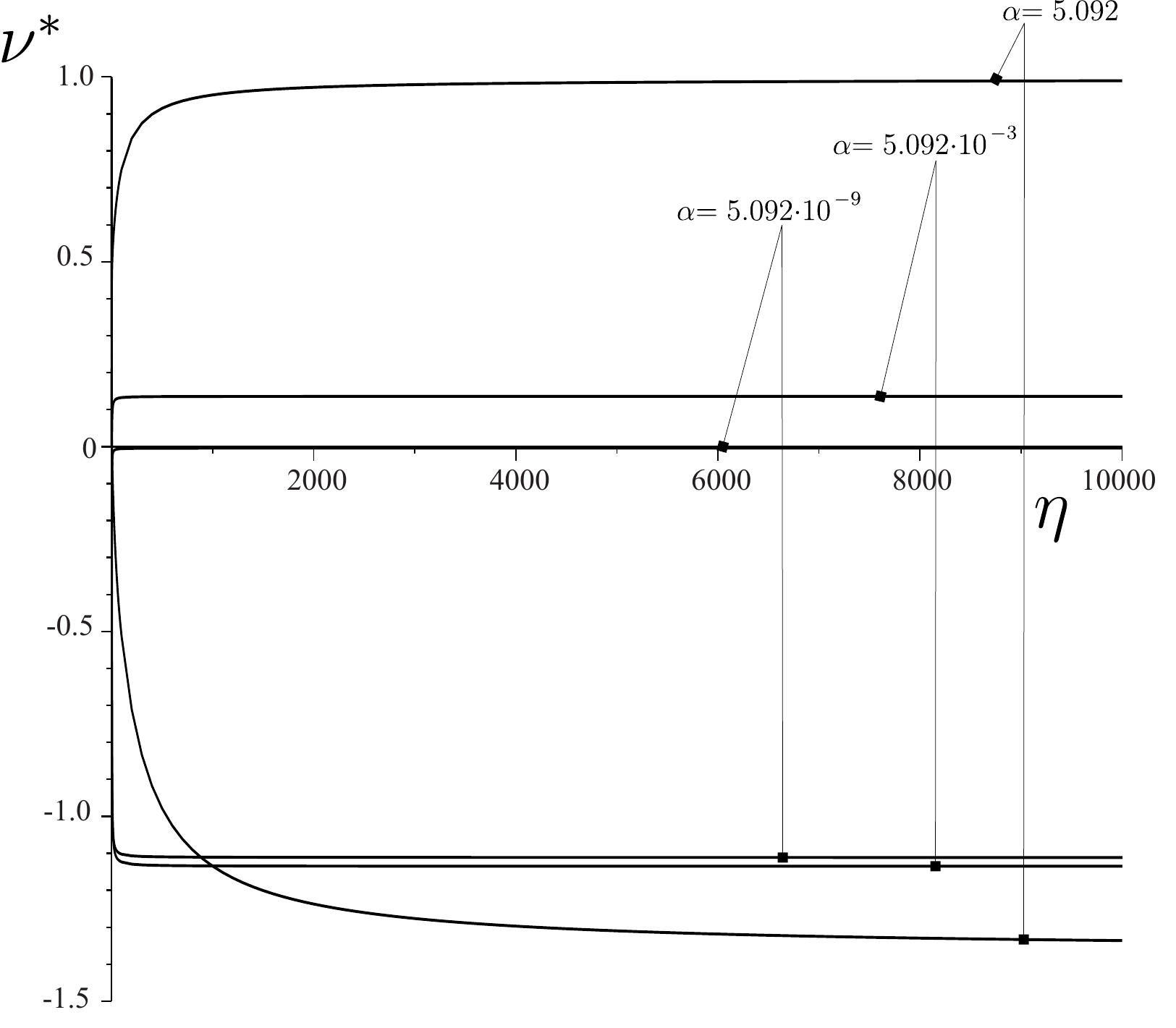}&
                    \includegraphics[height=4.5cm]{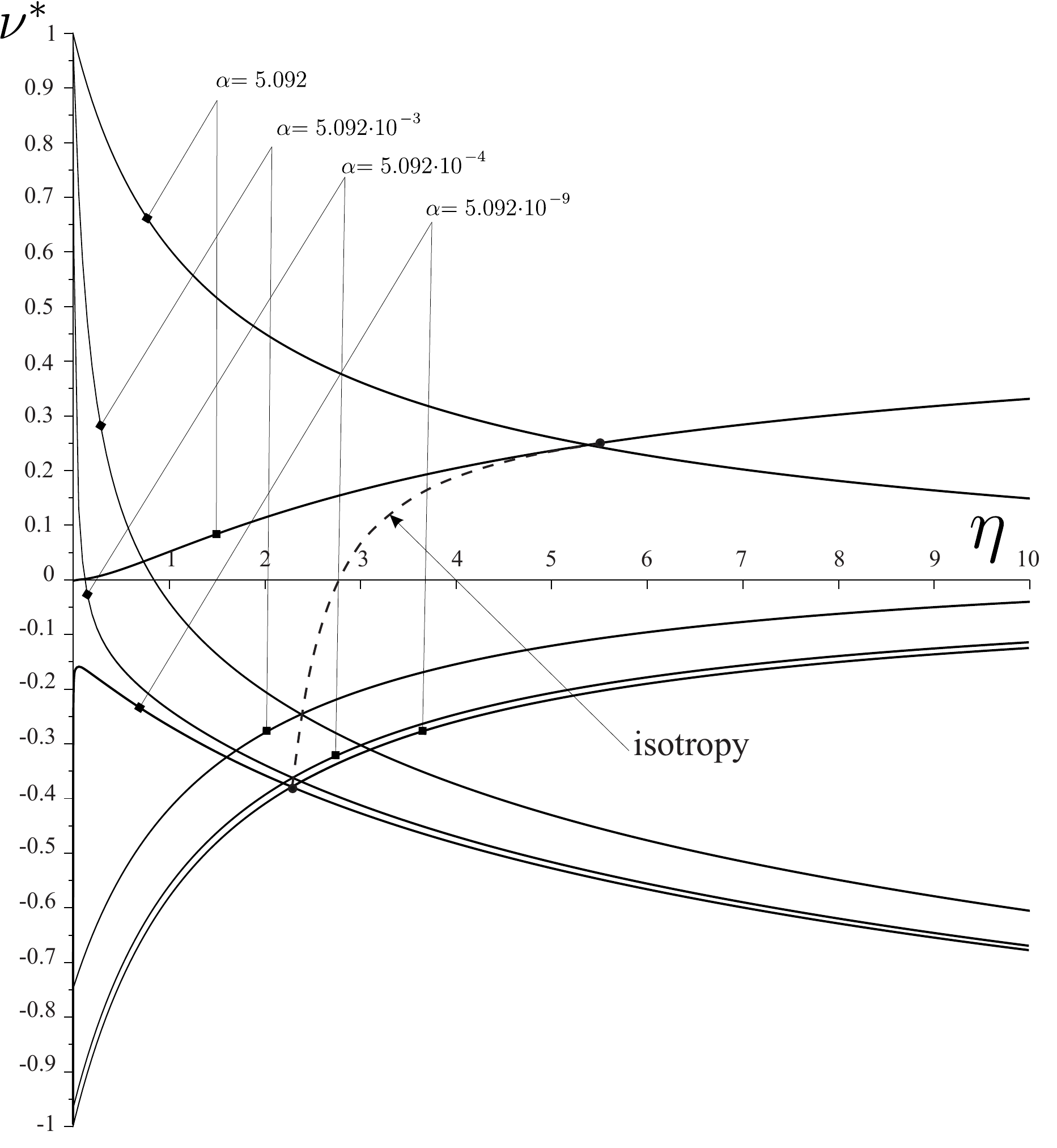}\\
         			(a) & (b)& (c)
         		\end{tabular} \\
                  \end{tabular}
       }
\caption{(a) Global minimum Poisson's ratio $\nu^*(\textbf{n},\textbf{m})$ for a cubic material in the $(\nu^*_{001},\nu^*_{1\bar{1}0})$ plane, for different values of the cross-section ratio $\eta$. (b)-(c) Global minimum and maximum Poisson's ratio $\nu^*$ for $\eta\in(1,10000)$ and $\eta\in(1,10)$, respectively. The dashed line indicates isotropic behavior.}
\label{fig2_2_70}
\end{figure}

We conclude the analysis of the lattice structure by investigating the behavior in the whole range of $\eta$ for different values of the ratio $\alpha$. The paths in the $(\nu^*_{001},\nu^*_{1\bar{1}0})$ plane are shown in Fig. \ref{fig2_2_70}a where the global minimum $\nu^*(\textbf{n},\textbf{m})$ in different regions is also indicated. In Figs. \ref{fig2_2_70}b and \ref{fig2_2_70}c we show $\nu^*_{min}$ and
$\nu^*_{max}$ as a function of $\eta$, for different values of $\alpha$.
Isotropic behavior is attained on the line $\nu^*_{001}=\nu^*_{1\bar{1}0}$, corresponding to the intersection points between $\nu^*_{min}$ and $\nu^*_{max}$ in Fig. \ref{fig2_2_70}c highlighted with a dashed line.

It is interesting to focus on the behaviour when $\eta\gg 1$. In such a case the minimum Poisson's ratio $\nu^*_{min}$ is $\nu^*_1<-1$, tending to the stability limit $\mu^*\to\infty$ on the line $\nu^*_{1\bar{1}0}=-1$ when $\eta\to \infty$. For $\alpha=5.092$ the limiting value is $\nu^*_1= -1.32$.
The maximum is  $-1<\nu^*_0\le 0$ for $\nu^*_{001}\le 0$,  $\nu^*_{001}$ in the interval $0\le\nu^*_{001}\le 1.5$ and $\nu^*_2>1.5$ for $1.5\le\nu^*_{001}\le 2$. Omni-directional negative Poisson's ratio is obtained for $\alpha\lesssim 10^{-4}$, namely $\nu^*_{max}=\nu^*_0$, while $\nu^*_{max}=\nu^*_{001}$ for  $\alpha\gtrsim 10^{-4}$.

Finally, it appears that increasing $\alpha$ the accumulation point for $\eta\to\infty$ is $(\nu^*_{001},\nu^*_{1\bar{1}0})=(1,-1)$, corresponding to $\nu^*_{min}=\nu^*(110.73,1\bar{1}0)=-(\sqrt{3}+1)/2=-1.366$ and  $\nu^*_{max}=\nu^*_{001}=1$.

\section{Conclusions}
\label{Sect04}

We have proposed a new family of three dimensional lattices with negative effective Poisson's ratio. The auxetic behavior is given by the topology of the microstructure and, in particular, by the contrast between the stiffnesses (longitudinal and flexural) of the elements of the microstructure.
For the cubic structure the constitutive stability domain is a triangle in the $(\nu^*_{001},\nu^*_{1\bar{1}0})$ plane and the minimum and maximum Poisson's ratios are achieved for infinite contrasts, when the boundary of the domain is reached asymptotically. The minimum is  attained for $\alpha\to 0$ and $\eta/\alpha\to\infty$ and maximum for $\alpha\to \infty$. It is shown that the lattice can be designed to have Poisson's ratio $-1$, a case of perfect dilational behavior. The structure can also be modified in order to have Poisson's ratio less than $-1$ along some oblique direction with a limiting value of $\nu^*=-(\sqrt{3}+1)/2$. In addition, an isotropic structure is proposed which can have omni-directional and uniform negative Poisson's ratio.

The microstructured medium is scale-independent, so that it is possible to think to technological applications ranging from large scale structural systems as in Civil and Aeronautical Engineering, where hinges are actually built, to micro- and nanostructures leading to metamaterials.

\section*{Acknowledgment}
M.B. and L.C. acknowledge the financial support of the Regione Autonoma della Sardegna (LR7 2010, grant `M4' CRP-27585). Finally, the authors would like to thank Prof. Andrew Norris for the discussion about Poisson's ratio in cubic materials.

\end{document}